\documentclass[aip,jmp,reprint,amsmath,amssymb,onecolumn]{revtex4-1}

\usepackage{bm}

\newcommand\lgr{\mathfrak{g}}\newcommand\lgl{\mathfrak{gl}}
\newcommand\lsl{\mathfrak{sl}}\newcommand\lo{\mathfrak{o}}
\newcommand\lsp{\mathfrak{sp}}\newcommand\lu{\mathfrak{u}}

\begin{document}

\title{Proof by characters of the orthogonal-orthogonal duality and
  relations of Casimir invariants}

\author{K. Neerg\aa rd}
\affiliation{Fjordtoften 17, 4700 N\ae stved, Denmark}

\begin{abstract}
  The theorem of orthogonal-orthogonal duality of Rowe, Repka, and
  Carvalho is proven by a method based on characters that is very
  different from theirs and akin to Helmers's proof from half a
  century earlier of the analogous sympletic-symplectic duality. I
  demonstrate how three duality theorems listed by Rowe, Repka, and
  Carvalho allow very brief derivations of linear relations between
  the Casimir invariants of the connected representations based on the
  geometry of their Young diagrams, and discuss for which physical
  systems other than such already considered in the literature an
  analysis in terms of the orthogonal-orthogonal duality might be
  useful.
\end{abstract}

\maketitle

\section{\label{sec:intr}Introduction}

In 1961 Helmers proved a remarkable theorem that establishes a 1--1
correspondence between the irreps of two commuting symplectic Lie
algebras of operators on the Fock space of several kinds of fermions
inhabiting a common 1-kind configuration space of even
dimension.~\cite{ref:Hel61} Helmers's article appeared in a context of
contemporary efforts~\cite{ref:Boh58,ref:Bog58,ref:Sol58a,*ref:Sol58b}
to adapt to nuclei the theory of superconductivity of Bardeen, Cooper,
and Schrieffer.~\cite{ref:Bar57a,*ref:Bar57b} In the simplest
application of his theorem, only one kind of nucleon, neutron or
proton, is considered. A basis for the 1-nucleon configuration space
may consist of one or several shells of orbits, where the orbits
within a shell share the value of the angular momentum quantum number
$j$, or it may be composed of pairs of time reversed stationary states
in a deformed potential well. One symplectic Lie algebra is then
induced by infinitesimal symplectic transformations of that space. Due
to the antisymmetry of many-fermion wave functions, only irreps
associated with 1-column Young diagrams occur. They are described by
the quantum number of seniority defined for atoms by
Racah~\cite{ref:Rac43} in 1943, and their carrier spaces are
eigenspaces of a particular ``pairing'' interaction which acts only in
those pairs of time reversed orbits whence Cooper pairs are built.

In this simplest case, the second symplectic Lie algebra was
introduced by Kerman in the same context of adaption of the
Bardeen-Cooper-Schrieffer theory to nuclei.~\cite{ref:Ker61} It
includes operators that create or annihilate pairs of nucleons. It was
not recognized at its conception as a symplectic Lie algebra because
its structure is that of $\lsp(2)$, which is isomorphic to $\lsl(2)$.
Accordingly, Kerman's Lie algebra is known to nuclear physicists as
the quasi-spin algebra. The 1--1 correspondence between seniority and
quasi-spin established by Helmers is at the core of an extensive
analysis of the nuclear shell model pursued most prominently by Talmi
and his coworkers.~\cite{ref:Tal93} The virtue of the correspondence
of seniority and quasi-spin at the base of this analysis lies in the
fact that the quasi-spin algebra connects states of equal seniority in
nuclei with different numbers of nucleons. This allows conclusions
about states of a given seniority of more complex nuclei to be drawn
by recourse to systems with fewer nucleons.

A note on notation: In the present article, $\lsl(d)$ is $A_\Omega$ in
Cartan's classification,~\cite{ref:Car94,ref:Rac51,ref:Jac62} with
$d = \Omega + 1$, and $\lgl(d)$ is $\lsl(d)$ extended by a commuting
1-dimensional Lie algebra. The Lie algebra $\lo(d)$ is $B_\Omega$ or
$D_\Omega$ with $d = 2 \Omega + 1$ and $d = 2 \Omega$, respectively,
and $\lsp(d)$ is $C_\Omega$ with $d = 2 \Omega$. Having quantum
mechanical applications in mind, I understand that the base field is
the field of complex numbers.

Already in 1952, Flowers generalized Racah's treatment of systems with
one kind of fermions to systems of neutrons and protons sharing a
configuration space.~\cite{ref:Flo52b} The states of such a system may
carry irreps of the Lie algebra of infinitesimal symplectic
transformations of the 1-kind configuration space associated with
2-column Young diagrams. Let $d$ be the dimension of the 1-kind
configuration space. To describe $\lsp(d)$ irreps associated with
Young diagrams with maximally 2 columns, Flowers introduces a second
quantum number besides seniority, the quantum number of reduced
isospin. By Helmers's theorem, each pair of seniority and reduced
isospin then corresponds 1--1 to an irrep of a number non-conserving
$\lsp(4)$ algebra which commutes with the $\lsp(d)$ algebra. This work
by Flowers was preceded by many studies where spatial states of the
nuclear system were classified by representations of $\lo(d)$
associated with Young diagrams with maximally 4 columns, corresponding
to the 4 dimensions of the space of spin and isospin of a nucleon.
Here, $d$ denotes the dimension of the 1-nucleon spatial configuration
space. See Refs.~\onlinecite{ref:Jah50,ref:Flo52a} and refs. therein.

Independently of Helmers, Flowers and Szpikowski introduced a few
years after his work a ``generalized quasi-spin'' algebra, which they
describe as $\lo(5)$, and which is, in fact, identical to Helmers's
$\lsp(4)$ algebra.~\cite{ref:Flo64a} (The isomophism of these two Lie
algebras is well known.~\cite{ref:Jac62,ref:Rac51}) Shortly
thereafter, and returning to a separation of the degrees of freedom of
a single nucleon into spatial ones on the one hand and spin and
isospin on the other, these authors proposed an $\lo(8)$ algebra of
``quasi-spin in $LS$ coupling''.~\cite{ref:Flo64b} Flowers and
Szpikowski calculated the spectra of pairing interactions in terms of
Casimir invariants of these semi-simple Lie algebras and found that
the resulting expressions were identical, upon a suitable association
of quantum numbers, to such obtained in previous analyses in terms of
representations of $\lsp(d)$ or $\lo(d)$.~\cite{ref:Edm52,ref:Bay60}
In the case of $\lo(5)$ and $\lsp(d)$, the association of quantum
numbers could have been derived from Helmers's theorem, had the
authors been aware or it at the time of writing. Both algebras
$\lo(5)$ and $\lo(8)$ have attracted much attention from the nuclear
physics community within the last two decades related to discussions
of the possibility of condensation of Cooper pairs built from pairs of
a neutron and a proton besides an established presence of condensates
of Cooper pairs built from pairs of nucleons of the same kind.~\cite{
  ref:Eng96,ref:Eng97,ref:Kot06,ref:Duk06,ref:Ler07,ref:Row07,
  ref:Dru13,ref:Dru14,ref:Mar18}

Helmers anticipates in Ref.~\onlinecite{ref:Hel61} that results
similar to his hold in $LS$ coupling, but half a century would pass
before a theorem analogous to his and pertaining to the separations of
fermionic degrees of freedom into such of motion in space and
additional quantum numbers would appear in the physics
literature.~\cite{ref:Row11} (A proof pertaining to the case of
$\lo(8)$ was published a few years earlier.~\cite{ref:Row07}) In
Ref.~\onlinecite{ref:Row11}, Rowe, Repka, and Carvalho both present a
new proof of Helmers's theorem and prove an analogous theorem that
involves orthogonal Lie algebras. I follow their terminology and name
the correspondences established by these theorems the
symplectic-symplectic and orthogonal-orthogonal dualities,
respectively, abbreviated $\lsp$-$\lsp$ and $\lo$-$\lo$. In both
cases, one of the Lie algebras generalizes the algebras $\lsp(d)$ and
$\lo(d)$ mentioned above. The other one is $\lsp(2 k)$ or $\lo(2 k)$,
where $k$ is the number of ``kinds'' of particles. To make the latter
concept concrete, in the $\lsp$-$\lsp$ duality, $k$ equals 1 for
systems of only electrons, only neutrons or only protons, and $k = 2$
for the system of neutrons and protons. In relation to the $\lo(8)$
algebra of Flowers and Szpikowski, $k$ equals 4 corresponding to the 4
linearly independent states of spin and isospin of a nucleon. The
algebra $\lsp(d)$ or $\lo(d)$ will be referred to as the number
conserving Lie algebra and the algebra $\lsp(2 k)$ or $\lo(2 k)$ as
the number non-conserving Lie algebra.

It should be specified at this point that duality in the sense of
Ref.~\onlinecite{ref:Row11} and the present article requires more than
commuting Lie algebras and a 1--1 correspondence of representations.
Each product of connected representations must also be realized with
multiplicity 1 on the total Hilbert space, which is the Fock space in
this case.

The proofs in Ref.~\onlinecite{ref:Row11} take recourse to yet another
duality, which is proven there, as well. The authors call it a
unitary-unitary duality, but the name of a $\lgl$-$\lgl$ duality would
be equally justified. It refers to an observation that is known and
applied since long ago in the theory of many-fermion systems: Let the
configuration space of one fermion be spanned by wave functions
$\chi(p)\psi(\tau)$, where $p$ and $\tau$ are sets of quantum numbers
such as spatial and spin quantum numbers. Let $\lgl(d)$, where $d$ is
the number of values of $p$, denote the tensor representation on the
space of functions $\phi(p_1,\dots,p_n,\tau_1,\dots,\tau_n)$ of the
algebra of infinitesimal linear transformations of the function
$\chi$, and $\lgl(k)$ similarly in terms of $\psi$ and the number $k$
of values of $\tau$. Then $\lgl(d)$ and $\lgl(k)$ are dual on the
space of $n$-fermion wave functions in the sense above, and the
connected irreps are associated with Young diagrams which result from
one another by the interchange of rows and columns. Such Young
diagrams are called conjugate. The assertion follows from the facts
that $\phi$ can be expanded on products
$\chi(p_1,\dots,p_n) \psi(\tau_1,\dots,\tau_n)$, and that, say,
$\lgl(d)$ and the symmetric group $S_n$ are dual on the space of
functions $\chi$.~\cite{ref:Wey39,ref:Lit40} (The Schur-Weyl duality
in the terminology of Ref.~\onlinecite{ref:Row11}.) Antisymmetrization
in the pairs $(p_i,\tau_i)$ leaves one irreducible
$\lgl(d) \oplus \lgl(k)$ module for each pair of conjugate Young
diagrams such that the corresponding $S_n$ irreps can be realized on
functions $\chi$ and $\psi$.

In Ref.~\onlinecite{ref:Row11} the proofs of the $\lsp$-$\lsp$ and
$\lo$-$\lo$ dualities proceed from the observation that the number
conserving Lie algebra is contained in a $\lgl(d)$ algebra and the
number non-conserving one contains a $\lgl(k)$ algebra. Each
irreducible module of their direct sum (more precisely in the
$\lo$-$\lo$ case, of the outer product $O(d) \otimes SO(2 k)$ of the
full and proper orthogonal groups) is then shown to contain a state
which has the highest weight with respect to irreps of both $\lgl$
algebras. Helmers's original proof of the $\lsp$-$\lsp$ duality is
very different from this and more direct. It uses the expression for
the $\lsp$ characters found early in the last
century.~\cite{ref:Wey39,ref:Lit40} It seems instructive to prove also
the $\lo$-$\lo$ duality in this manner. This is done in
Secs.~\ref{sec:or}--\ref{sec:prf} of the present article, where the
theorem is also stated more symmetrically in $\lo(d)$ and $\lo(2 k)$
than in Ref.~\onlinecite{ref:Row11}.

In the applications to nuclear spectroscopy mentioned above, Casimir
invariants play a central role. The Casimir invariants of dually
connected irreps must be related, and these relations turn out to be
linear. Some of these relations were derived algebraically, and such
derivations were seen occasionally as proofs of the duality, which of
course they are not.~\cite{ref:Hel61,ref:Kot06,ref:Dru13,ref:Dru14} It
amounts to essentially the same that in several cases the spectra of
pairing interactions where expressed by Casimir invariants of either
one of a pair of dual Lie algebras and the expressions obtained were
found to agree upon suitable associations of quantum
numbers.~\cite{ref:Tal93,ref:Flo64a,ref:Flo64b} (In the particular
case of the $\lsp(d)$--$\lsp(2)$ duality,~\cite{ref:Tal93} the 1--1
correspondence actually follows from this comparison, but multiplicity
1 of the products of representations does not.) In
Secs.~\ref{sec:spgl}--\ref{sec:cas}, I show how to derive in a very
simple manner based on the geometry of the Young diagrams the linear
relations of Casimir invariants directly from the associations of
representation according to the $\lo$-$\lo$, $\lsp$-$\lsp$, and
$\lgl$-$\lgl$ duality theorems. A more complicated, algebraic
derivation akin to those of Refs.~\onlinecite{ref:Hel61} and
\onlinecite{ref:Kot06} is given in one case for comparison.

Before summarizing the article in Sec.~\ref{sec:sum}, I address in
Sec.~\ref{sec:phys} the question of possible applications of the
orthogonal-orthogonal duality to the analysis of actual physical
systems other than such already considered in the literature.

\section{\label{sec:or}Orthogonal Lie algebras}

To prepare the proof of the $\lo$-$\lo$ duality in the manner of
Helmers, I describe in this section the construction of the two
commuting $\lo$ algebras. A number $k$ of kinds of fermions is
considered, and the kind is denoted by letters
$\tau,\upsilon,\dots$\,. These fermions inhabit a common 1-kind
configuration space of dimension $d$ with orthonormal basic states
denoted by $| p \rangle, | q \rangle, \dots$\,. The annihilator of a
fermion of kind $\tau$ in the state $|p \rangle$ is denoted by
$a_{p\tau}$. By definition, the Fock space $\Phi$ is spanned by the
states generated from the vacuum by the operators
\begin{equation}
   1 , \quad a^\dagger_{p\tau} , 
   \quad a^\dagger_{p\tau} a^\dagger_{q\upsilon} , \quad \dots \quad .
\end{equation}

\subsection{\label{sec:cons}Number conserving Lie algebra}

A Lie algebra of generators $x$ of orthogonal transformations of the 1-kind
configuration space is defined by the conservation of a non-singular
symmetric tensor $\langle p q | g \rangle = \langle q p | g \rangle$
in the sense that
\begin{equation}\label{eq:cons_s}
  \sum_r ( \langle p | x | r \rangle \langle r q | g \rangle
    + \langle q | x | r \rangle \langle p r | g \rangle ) = 0 .
\end{equation}
It will be assumed that
\begin{equation}\label{eq:g_diag-s}
  \langle p q | g \rangle = \delta_{pq}
\end{equation}
for some choice of the states $| p \rangle$. The matrix
$\langle p | x | q \rangle$ then is antisymmetric. If the 1-kind
configuration space carries integral angular momentum quantum numbers
$l$, one can write $| p \rangle = | \alpha l m \rangle$, where $m$ is
the magnetic quantum number. In atomic and nuclear physics, $g$ is
then often taken to be the Wigner metric~\cite{ref:Edm57}
\begin{equation}~\label{eq:Wig-l}
  \langle \alpha l m , \alpha' l' m'| g \rangle
    = (-)^{l+m} \delta_{\alpha\alpha'} \delta_{ll'} \delta_{m,-m'} .
\end{equation}
Because this can be given the form~\eqref{eq:g_diag-s} by a unitary
transformation, it is encompassed by the assumption above. Using the
metric~\eqref{eq:Wig-l} instead of \eqref{eq:g_diag-s} only adds
complication.

Operators $x_{pq}$ can be defined by
\begin{equation}
  x_{pq} | q \rangle = | p \rangle, \quad
  x_{pq} | p \rangle = - | q \rangle, \quad
  x_{pq} | r \rangle = 0, \quad
  x_{pp} = 0,  \quad p,q,r \text{ different} .
\end{equation}
They satisfy $x_{pq} = - x_{qp}$, and those with $p > q$ form a basis
for the Lie algebra. The index $p$ of the basic 1-kind states states
$|p \rangle$ may be assumed to take the values $-\Omega, -\Omega + 1,$
$\ldots, \Omega$ with 0 omitted when $d$ is even. The operators
$h_p = -i x_{p,-p},\ p > 0,$ then form a basis for a Cartan subalgebra
which gives the roots of Ref.~\onlinecite{ref:Rac51}. Because the root
diagram is invariant under reflections and permutations of the
coordinate axes, the Killing form $(x,x')$ can be so renormalized that
$(h_p,h_q) = \delta_{pq}$. I here include for convenience the case
$d = 2$ when the Killing form vanishes because the Lie algebra is
1-dimensional and therefore Abelian. For $d = 1$ the Lie algebra is
0-dimensional (has one element, 0) so the Killing form is undefined.

For every root $\rho$ an operator $x_\rho$ is determined within a
normalization by
\begin{equation}
  [ h_p , x_\rho ] = \rho_p x_\rho ,
\end{equation}
and the operators $h_p$ and $x_\rho$ form a basis for the Lie algebra.
The entire Killing form is then determined by the
relations~\cite{ref:Jac62,ref:Rac51}
\begin{equation}
  ( h_p , x_\rho ) = 0 , \quad
  ( x_\rho , x_\sigma ) = 0, \quad \sigma \ne - \rho , \quad
  [ x_\rho , x_{-\rho} ]
  = ( x_\rho , x_{-\rho} ) \sum_p \rho_p h_p .
\end{equation}
Transformation to the basis of operators $x_{pq}$ gives the simple
result
\begin{equation}\label{eq:Kil}
  ( x_{pq} , x_{rs} ) = \delta_{pq,sr} 
\end{equation}

The basic operators $h_p$ of the Cartan subalgebra are diagonalized by
the unitary transformation to basic states
$| \nu \rangle, | \pi \rangle, \dots$ defined by~\cite{ref:Wey39}
\begin{equation}\label{eq:trans}
  | \pm \nu \rangle = \sqrt{\tfrac12}
    ( | p \rangle \pm i | - p \rangle ) , \quad \nu = p > 0, \quad
  | \nu \rangle = | p \rangle , \quad \nu = p = 0 .
\end{equation}
In fact,
\begin{equation}\label{eq:nu}
  h_\nu | \pm \nu \rangle = \pm | \nu \rangle, \quad
  h_\nu | \pi \rangle = 0, \quad \pi \ne \pm \nu .
\end{equation}

The Lie algebra of operators $x$ is faithfully represented on $\Phi$
by the operators
\begin{equation}\label{eq:S}
  X = \sum_{pq\tau} \langle p | x | q \rangle
    a^\dagger_{p\tau} a_{q\tau} .
\end{equation}
The representative of $h_p$ will be denoted by $H_p$. Evidently every
$X$ commutes with the total number
\begin{equation}
  n = \sum_{p\tau} a_{p\tau}^\dagger a_{p\tau}
\end{equation}
of fermions. From now on, the symbol $\lo(d)$ is reserved for the
orthogonal Lie algebra of these operators $X$. By Eq.~\eqref{eq:Kil},
its Casimir operator $C_\text{$\lo(d)$}$ is given by
\begin{equation}\label{eq:Co-d-o}
  C_\text{$\lo(d)$}
    = \sum_{p > q} X_{pq} X_{qp} = \tfrac12 \sum_{pq} X_{pq}  X_{qp} .
\end{equation}

\subsection{\label{sec:ncons}Number non-conserving Lie algebra}

I turn to the construction of the number non-conserving orthononal Lie
algebra. To this end let $\phi,\chi,\dots$ denote arbitrary linear
combinations of the fermion fields $a_{p\tau}$ and
$a^\dagger_{p\tau}$, and let $Y$ be any linear combination of
commutators $[\phi,\chi]$. The set of such $Y$ is closed under
commutation.

One may now inquire which $Y$ commute with every $X$. Considering
first operators
\begin{equation}\label{eq:dag-none}
  Y =\tfrac12 \sum_{pq\tau\upsilon}
    \langle p \tau | y | q \upsilon \rangle
      [ a^\dagger_{p\tau} , a_{q\upsilon} ]
    = \sum_{pq\tau\upsilon} \langle p \tau | y | q \upsilon \rangle
      ( a^\dagger_{p\tau} a_{q\upsilon}
      - \tfrac12 \delta_{p\tau,q\upsilon} ) ,
\end{equation}
one finds that each matrix in $p$ and $q$ with elements
$\langle p \tau | y | q \upsilon \rangle$ must commute with every
matrix $\langle p | x | q \rangle$. For $d \ne 2$ the representation
of the Lie algebra of operators $x$ by itself is irreducible. (Cf.
Sec.~\ref{sec:rep}. For $d > 1$ this representation is described by
the 1-cell Young diagram. For $d = 1$ the 1-kind space is trivially
irreducible.) By Schur's lemma, each matrix
$\langle p \tau | y | q \upsilon \rangle$ is then proportional to the
unit matrix, so the space of operators $Y$ of the
form~\eqref{eq:dag-none} that commute with every $X$ is spanned by the
operators
\begin{equation}\label{eq:Y+-}
  Y_{\tau,-\upsilon} := - Y_{-\upsilon,\tau}
  = \sum_p a^\dagger_{p\tau} a_{p\upsilon}
    - \frac d 2 \delta_{\tau\upsilon} .
\end{equation}

As to operators
\begin{equation}\label{eq:dag-dag}
  Y =\sum_{pq\tau\upsilon} \langle p \tau , q \upsilon | y \rangle
    a^\dagger_{p\tau} a^\dagger_{q\upsilon} ,
\end{equation}
the requirement is
\begin{equation}
  \sum_r ( \langle p | x | r \rangle
    \langle r \tau , q \upsilon | y \rangle
    + \langle q | x | r \rangle
    \langle p \tau , r \upsilon | y \rangle ) = 0 .
\end{equation}
By the antisymmetry of the matrix $\langle p | x | y \rangle$ the
matrix in $p$ and $q$ with elements
$\langle p \tau , q \upsilon | y \rangle$ then commutes with
$\langle p | x | q \rangle$. For $d \ne 2$ it follows again that
$\langle p \tau , q \upsilon | y \rangle$ is proportional to
$\delta_{pq}$. The space of operators $Y$ of the
form~\eqref{eq:dag-dag} that commute with every $X$ is then spanned by
the operators
\begin{equation}\label{eq:dd-bas}
  Y_{\tau\upsilon}
    = \sum_p a^\dagger_{p\tau} a^\dagger_{p\upsilon} .
\end{equation}
The hermitian conjugates of the operators~\eqref{eq:dd-bas} will be
denoted by $Y_{-\upsilon,-\tau}$ and span the space of linear
combinations of products of pairs of annihilators that commute with
every $X$.

All the basic operators defined above have the form
\begin{equation}
 Y_{\alpha\beta}\label{Y-bas}
    = \tfrac12 \sum_p [ a^\dagger_{p\alpha} , a^\dagger_{p\beta} ] ,
\end{equation}
where $\alpha$ and $\beta$ take values $\pm 1, \pm 2, \dots , \pm k$,
and $a_{p,-\tau} := a^\dagger_{p\tau}$. One easily derives
\begin{equation}\label{eq:com}
  [ Y_{\alpha\beta} , Y_{\gamma\delta} ]
    = \delta_{\beta,-\gamma} Y_{\alpha\delta}
      - \delta_{\beta,-\delta} Y_{\alpha\gamma}
      - \delta_{\alpha,-\gamma} Y_{\beta\delta}
      + \delta_{\alpha,-\delta} Y_{\beta\gamma} .
\end{equation}
By comparison with Ref.~\onlinecite{ref:Rac51} it follows that the
span of the set of operators~\eqref{Y-bas} is an $\lo(2 k)$ algebra,
and the symbol $\lo(2 k)$ is reserved from now on for this orthogonal
Lie algebra. The operators $Y_{\tau,-\tau}$ form a basis for a Cartan
subalgebra which gives the roots of Ref.~\onlinecite{ref:Rac51}.
Renormalizing the Killing form $(Y,Y')$ so that these operators have
squared Killing norm 1 results in
\begin{equation}
  ( Y_{\alpha\beta} , Y_{\gamma\delta} ) =
    \delta_{\alpha\beta;-\delta,-\gamma} .
\end{equation}
The Casimir operator $C_\text{$\lo(2 k)$}$ is then given by
\begin{equation}\label{eq:Co-2k}
  C_\text{$\lo(2 k)$} = \sum_{\alpha>\beta} Y_{\alpha\beta} Y_{-\beta,-\alpha}
    = \tfrac12 \sum_{\alpha\beta} Y_{\alpha\beta} Y_{-\beta,-\alpha} .
\end{equation}

For $d = 2$ the algebra $\lo(2 k)$ is not the maximal set of operators
$Y$ that commute with every $X$. The following discussion holds,
anyway, in this case, as well. (For $d = 2$ the maximal set forms a
$\lgl(2 k)$ algebra and contains $\lo(d)$. If also $k > 1$, the Lie
algebra $\lo(d)$ is the maximal subalgebra of this $\lgl(2 k)$ algebra
that commutes with $\lo(2 k)$. If $k = 1$, both Lie algebras are
1-dimensional, and they span a 2-dimensional Abelian Lie algebra.)

The unitary transformation that transforms the
metric~\eqref{eq:g_diag-s} to a general symmetric $g$ transforms the
operators~\eqref{eq:dd-bas} to
\begin{equation}\label{eq:dd-bas-g}
  Y_{\tau\upsilon}
    = \sum_{pq} a^\dagger_{p\tau} a^\dagger_{q\upsilon}
      \langle p q | g \rangle .
\end{equation}
The form of the operators ~\eqref{eq:Y+-} does not change by this
transformation.

\section{\label{sec:thrm}Orthogonal-orthogonal duality}

\subsection{\label{sec:rep}Representations of orthogonal Lie algebras}

Before turning to the proof of the duality theorem it is necessary
to recapitulate some facts of representations of orthogonal Lie
algebras. My notation refers to $\lo(d)$, but the discussion applies
analogously to $\lo(2 k)$ upon evident substitutions. I shall
similarly not in the subsequent sections state explicitly results for
$\lo(2 k)$, $\lsp(2 k)$, or $\lgl(k)$ which follow by analogy from
those for $\lo(d)$, $\lsp(d)$, or $\lgl(d)$.

The Lie algebra $\lo(d)$ has representations described by
\textit{partitions} $[\lambda]$, which are sequences of \textit{parts}
$\lambda_1,\lambda_2,\dots,\lambda_\Omega$ with
\begin{equation}\label{eq:part}
  0 \le \lambda_1 \le \lambda_2 \le \dots \le \lambda_\Omega .
\end{equation}
(See Refs.~\onlinecite{ref:Wey39,ref:Lit40}, noticing that the Lie
algebra generates the proper orthogonal group.) Unconventionally, I
number the parts in non-decreasing order, which will turn out
convenient. The convention is to number them with the largest first.
Either all $\lambda_p$ are integral or all of them are half-integral.
Representations with half-integral parts are spin representations. The
representation described by the partition $[\lambda]$ is irreducible
except when $d$ is even and $\lambda_1 > 0$. When this happens, the
representation splits into two inequivalent irreps. They may be
distinguished by opposite signs of $\lambda_1$, in which case
$|\lambda_1|$ replaces $\lambda_1$ in the
inequalities~\eqref{eq:part}. Then $\lambda_p$ is always the
eigenvalue of $H_p$ on the vector of highest weight of an irreducible
module.\cite{ref:Boe63} I consider here only non-negative $\lambda_1$
and understand that $[\lambda]$ then describes the total, reducible
representation when $d$ is even and $\lambda_1 > 0$. By definition,
the empty partition describes a 1-dimensional representation by the
constant 0.

A partition is visualized by a Young diagram, whose rows have lengths
$\lambda_p$ and are stacked from bottom to top in the order of $p$. In
particular the Young diagram of a spin representation thus has a
leftmost column of width $\frac12$ and height $\Omega$.

In the theory of characters a central role is played by the
numbers
\begin{equation}\label{eq:l-e}
  l_p = \lambda_p + p - 1
\end{equation}
when $d$ is \textit{even}, and
\begin{equation}\label{eq:l-o}
  l_p = \lambda_p + p - \tfrac12
\end{equation}
when $d$ is \textit{odd}. The number $2(l_p - \lambda_p)$ is the $p$th
component of the sum of positive roots. I call the strictly increasing
sequence of the numbers $l_p$ associated with a given partition
$[\lambda]$ the \textit{unfolded} partition and denote it by $[[l]]$.

\subsection{\label{sec:stmnt}Theorem}

The theorem of $\lo$-$\lo$ duality can now be stated as follows.

\smallskip

\itshape

Theorem: The Fock space $\Phi$ has the decomposition
\begin{equation}\label{eq:sum}
   \Phi = \bigoplus \textup{X}_{[\lambda]} \otimes \Psi_{[\mu]} ,
\end{equation}
where $\textup{X}_{[\lambda]}$ and $\Psi_{[\mu]}$ carry single
representations of \textup{$\lo(d)$} and \textup{$\lo(2 k)$}, respectively,
described by the partitions $[\lambda]$ and $[\mu]$. In the sum, each
pair of $[\lambda]$ and $[\mu]$ that satisfies the following criterion
appears exactly once. Let $[[l]]$ and $[[m]]$ be the corresponding
unfolded partitions, and consider the set of numbers that appear in
either $[[l]]$ or $[[m]]$. The criterion is that this set consists of
the numbers
\begin{equation}\label{eq:seq-e}
  0, 1, ... \, , \Omega + k - 1
\end{equation}
when $d$ is even, and
\begin{equation}\label{eq:seq-o}
  \tfrac12, \tfrac32, ... \, , \Omega + k - \tfrac12
\end{equation}
when $d$ is odd.

\upshape

\smallskip

It may be noticed that when $d$ is odd, so as when the 1-kind
space is a single $l$ shell, the $\lo(2 k)$ representations are spin
representations. The $\lo(d)$ representations are always non-spin
representations.

The connection between the partitions $[\lambda]$ and $[\mu]$ is
analogous to that of the Helmers theorem for symplectic Lie
algebras,~\cite{ref:Hel61} and can, like the latter, be expressed in
geometric terms: Consider a rectangle of width $k$ and height $d/2$.
In this rectangle, place the $[\lambda]$ Young diagram at the upper
left corner and the $[\mu]$ Young diagram at the lower right corner,
rotated 180$^\circ$ and reflected in the bisector of its right angle.
Then the two diagrams must fill the rectangle without overlap. See
Fig.~1 of Ref.~\onlinecite{ref:Hel61} for an illustration. It follows
that $[\lambda]$ has no part greater than $k$, and $[\mu]$ has no part
greater than $d/2$. Proving that this criterion is equivalent to that
of the theorem goes as in the symplectic case with a very minor
modification due to a different relation corresponding to the
relations~\eqref{eq:l-e} and \eqref{eq:l-o} and a different sequence
corresponding to the sequences~\eqref{eq:seq-e} and \eqref{eq:seq-o},
cf. Sec.~\ref{sec:sp}. I leave it to any interested reader to work it
out based on Ref.~\onlinecite{ref:Hel61}.

Yet another formulation is that for each $p$ the $p$th row of the
$[\lambda]$ diagram with the numbering above and the $p$th column from
the left of the $[\mu]$ diagram, not counting a possible half-width
column, have total length $k$, or, equivalently, that for each $\tau$
the $\tau$th row of the $[\mu]$ diagram with the analogous numbering
and the $\tau$th column from the left of the $[\lambda]$ diagram have
total length $d/2$.

\section{\label{sec:prf}Proof of the theorem}

By Eq.~\eqref{eq:nu} the general member of the $\lo(d)$ Cartan
subalgebra can be written
\begin{equation}
  H = \sum_{\nu>0} \phi_\nu H_\nu
    = \sum_{\nu>0,\tau} \phi_\nu 
      ( a^\dagger_{\nu\tau} a_{\nu\tau}
        - a^\dagger_{-\nu,\tau} a_{-\nu,\tau} ) .
\end{equation}
The trace
\begin{equation}
   \chi_{[\lambda]} = \text{Tr} \, O_\text{$\lo(d)$} ,
\end{equation}
over the carrier space of the representation $[\lambda]$ of
\begin{equation}
  O_\text{$\lo(d)$} = \exp H = \prod_{\nu>0,\tau} \epsilon_\nu 
    ^{ \ a^\dagger_{\nu\tau}a_{\nu\tau}
      - a^\dagger_{-\nu,\tau}a_{-\nu,\tau} } ,
\end{equation}
where
\begin{equation}
  \epsilon_\nu = \exp \phi_\nu ,
\end{equation}
is a polynomial in $\epsilon_\nu^{\;\frac12}$ which specifies the
representation uniquely. It is known as the character of the
representation.~\cite{ref:Jac62} Similarly the trace
\begin{equation}
   \chi_{[\mu]} = \text{Tr} \, O_\text{$\lo(2 k)$}
\end{equation}
over the carrier space of the representation $[\mu]$, where
\begin{equation}
  O_\text{$\lo(2 k)$} =\exp \sum_\tau \psi_\tau Y_{\tau,-\tau}
  = \exp \sum_{\nu\tau} \psi_\tau 
   ( a^\dagger_{\nu\tau}a_{\nu\tau} - \tfrac12 )
  = \prod_{\nu\tau} \eta_\tau 
    ^{ \ a^\dagger_{\nu\tau}a_{\nu\tau} - \frac12 }
\end{equation}
with
\begin{equation}
  \eta_\tau = \exp \psi_\tau ,
\end{equation}
is a character of that representation.

Because $\lo(d)$ and $\lo(2 k)$ commute, the direct sum of their
Cartan subalgebras is a Cartan subalgebra of their direct sum, so the
character $\chi$ of a representation of $\text{$\lo(d)$} \oplus
\text{$\lo(2 k)$}$ is given by
\begin{equation}
  \chi = \text{Tr} \, O_\text{$\lo(d)$} O_\text{$\lo(2 k)$} .
\end{equation}
It is a polynomial in $\epsilon_\nu^{\;\frac12}$ and
$\eta_\tau^{\;\frac12}$. When $\chi$ is evaluated on $\Phi$, proving
the theorem requires the verification of the identity
\begin{equation}\label{eq:charsum}
  \chi = \sum \chi_{[\lambda]} \chi_{[\mu]} ,
\end{equation}
where the sum runs over the same combinations of $[\lambda]$ and
$[\mu]$ as in Eq.~\eqref{eq:sum}.

When $d$ is even and $\lambda_1 > 0$, the character $\chi_{[\lambda]}$
given by the expressions~\eqref{eq:chi-e-d} and \eqref{eq:chi-o-d}
below is the sum of characters of the two inequivalent irreps which
compose the representation $[\lambda]$.~\cite{ref:Lit40} The proof
then asserts that both these irreps are present together. Similarly
when $\mu_1 > 0$.

The verification of the identity~\eqref{eq:charsum} proceeds
separately for even and odd $d$. Before entering these separate cases,
I make a definition. The symbol
\begin{equation}
  | f_1(z) , f_2(z) , \dots , f_n(z) | ,
\end{equation}
where $f_1, \dots, f_n$ are any functions, denotes the determinant
whose rows have the form shown for $n$ different values of $z$
specified in the context

\subsection{\label{sec:even}Even $d$}

The verification of the identity~\eqref{eq:charsum} in the case of
even $d$ follows closely Helmer's proof of the $\lsp$-$\lsp$
duality.~\cite{ref:Hel61} It is noticed that $\Phi$ is the direct
product of 2-dimensional spaces, one for each pair of $\nu$ and
$\tau$. Basic states of each 2-dimensional space correspond to the
state $| \nu \tau \rangle$ of 1 fermion being empty or occupied. Each
such space contributes to $\chi$ a factor

\begin{equation}
  \eta_\tau^{-\frac12}
    + \epsilon_{|\nu|}^{\ \text{sgn} \, \nu} \eta_\tau^{\ \frac12} .
\end{equation}
The pair of states $| \pm \nu , \tau \rangle$, where $\nu > 0$, then
gives the factor
\begin{equation}
  ( \eta_\tau^{-\frac12}
    + \epsilon_{\nu} \eta_\tau^{\ \frac12}  )
  ( \eta_\tau^{-\frac12}
    + \epsilon_{\nu}^{-1} \eta_\tau^{\ \frac12} )
  = \epsilon_{\nu} + \epsilon_{\nu}^{-1}
    + \eta_\tau + \eta_\tau^{-1}
  = c_1(\epsilon_{\nu}) + c_1(\eta_\tau) ,
\end{equation} 
where
\begin{equation}
  c_\alpha(z) = \begin{cases}
    z^\alpha + z^{-\alpha} , & \alpha > 0 , \\
    1 , & \alpha = 0 .
  \end{cases}
\end{equation}
One arrives at
\begin{equation}
  \chi = \prod_{\nu > 0} F(\epsilon_\nu)
\end{equation}
with
\begin{equation}
  F(\epsilon) = \prod_\tau ( c_1(\epsilon) + c_1(\eta_\tau) ) .
\end{equation}

Evidently $F(\epsilon) = 0$ for $\epsilon = - \eta_\tau$.
Using
\begin{equation}
  | c_0(\eta) , c_1(\eta) , \dots , c_{k-1}(\eta) |
  = (-1)^{ 0 + 1 + \dots + (k-1) }
  | c_0(-\eta) , c_1(-\eta) , \dots , c_{k-1}(-\eta) | ,
\end{equation}
one gets
\begin{multline}\label{eq:a}
  N := (-1)^{ 0 + 1 + \dots + (k-1) }
    | c_0(\epsilon) , c_1(\epsilon) , \dots , 
      c_{\Omega-1}(\epsilon) | 
    | c_0(\eta) , c_1(\eta) , \dots , c_{k-1}(\eta) | \, \chi \\
    = | c_0(\epsilon) , c_1(\epsilon) , \dots
      c_{k-1}(\epsilon) ,
        c_0(\epsilon) F(\epsilon) , c_1(\epsilon) F(\epsilon) , 
        \dots , c_{\Omega-1}(\epsilon)  F(\epsilon) | .
\end{multline}
Here, in the first determinant, $\epsilon$ takes the values
$\epsilon_\nu$, and in the second one, $\eta$ takes the values
$\eta_\tau$. In the third determinant, $\epsilon$ takes the values
$-\eta_\tau$ in the first $k$ rows, and $\epsilon_\nu$ in the
remaining $\Omega$ rows.

Now, using
\begin{equation}
  c_n(\epsilon) F(\epsilon) = c_{n+k}(\epsilon)
    \text{$+$ terms proportional to $c_m(\epsilon)$ with
      $m < n + k$ ,}
\end{equation}
one gets
\begin{multline}
  N = | c_0(\epsilon) , c_1(\epsilon) , \dots , 
        c_{\Omega+k-1}(\epsilon) | \\
    = \sum (-1)^{ (m_1-0) + (m_2-1) + \dots + (m_k-(k-1)) }
      | c_{l_1}(\epsilon) , c_{l_2}(\epsilon) , \dots , 
        c_{l_{\Omega}}(\epsilon) | 
      | c_{m_1}(-\eta) , c_{m_2}(-\eta) , \dots , 
        c_{m_k}(-\eta) | \\ 
    = (-1)^{ 0 + 1 + \dots + (k-1) } \sum
      | c_{l_1}(\epsilon) , c_{l_2}(\epsilon) , \dots , 
        c_{l_{\Omega}}(\epsilon) |
      | c_{m_1}(\eta) , c_{m_2}(\eta) , \dots ,
        c_{m_k}(\eta) | , 
\end{multline}
where the sum runs over all the pairs of unfolded partitions $[[l]]$
and $[[m]]$ of the theorem, and $\epsilon$ and $\eta$ take the values
$\epsilon_\nu$ and $\eta_\tau$, respectively. The
identity~\eqref{eq:charsum} now follows by~\cite{ref:Wey39,ref:Lit40}
\begin{equation}\label{eq:chi-e-d}
  \chi_{[\lambda]} = \frac
    { | c_{l_1}(\epsilon) , c_{l_2}(\epsilon) , \dots , 
        c_{l_{\Omega}}(\epsilon) | }
    { | c_0(\epsilon) , c_1(\epsilon) , \dots 
      c_{\Omega-1}(\epsilon) | }
\end{equation}
and the analogous expression for $\chi_{[\mu]}$ (written explicitly in
Eq.~\eqref{eq:chi-o-d}).

\subsection{\label{sec:odd}Odd $d$}

The idea of the verification of the identity~\eqref{eq:charsum} for
odd $d$ is the same, but the details are slightly more involved. There
is now a state $| \nu \rangle$ with $\nu = 0$, which contributes to
$\chi$ an extra factor
\begin{equation}
  f = \prod_\tau c_{\frac12} (\eta_\tau) .
\end{equation}
This factor can be combined with
$| c_0(\eta) , c_1(\eta) , \dots , c_{k-1}(\eta) |$ to give
\begin{multline}
   | c_0(\eta) , c_1(\eta) , \dots , c_{k-1}(\eta) | f
   = | c_{\frac12}(\eta) , c_{\frac32}(\eta) , \dots ,
       c_{k-\frac12}(\eta) | \\
   = (-1)^{ - \frac12 - \frac32 - \dots - (k - \frac12) }
     | s_{\frac12}(-\eta) , s_{\frac32}(-\eta) , \dots ,
       s_{k-\frac12}(-\eta) | ,
\end{multline}
where
\begin{equation}
  s_\alpha(z) = z^\alpha - z^{-\alpha} ,
\end{equation}
and $(-z)^\alpha$ is shorthand for $\exp (i \alpha \pi) z^\alpha$. One gets
\begin{multline}
  N := (-1)^{ \frac12 + \frac32 + \dots + (k - \frac12) }
    | s_{\frac12}(\epsilon) , s_{\frac32}(\epsilon) , \dots , 
      s_{\Omega-\frac12}(\epsilon) |
    | c_0(\eta) , c_1(\eta) , \dots ,
      c_{k-1}(\eta) | \, \chi \\
    = | s_{\frac12}(\epsilon) , s_{\frac32}(\epsilon) , \dots
      s_{k-\frac12} ,
        s_{\frac12}(\epsilon) F(\epsilon) ,s_{\frac32}(\epsilon) F(\epsilon) , 
        \dots , s_{\Omega-\frac12}(\epsilon)  F(\epsilon) |
\end{multline}
with the same $\epsilon$ and $\eta$ as in Eq.~\eqref{eq:a}.

Using
\begin{equation}
  s_\alpha(\epsilon) F(\epsilon) = s_{\alpha+k}(\epsilon)
    \text{$+$ terms proportional to $s_\beta(\epsilon)$ with
      $\beta < \alpha + k$}
\end{equation}
for half-integral $\alpha$ and $\beta$, one obtains
\begin{multline}
  N = | s_{\frac12}(\epsilon) , s_{\frac32}(\epsilon) , \dots
        s_{\Omega+k-\frac12} | \\
    = \sum (-1)^{ (m_1-\frac12) + (m_2-\frac32) + \dots
        + (m_k-(k-\frac12)) }
      | s_{l_1}(\epsilon) , s_{l_2}(\epsilon) , \dots , 
        s_{l_{\Omega}}(\epsilon) |
      | s_{m_1}(-\eta) , s_{m_2}(-\eta) , \dots , 
        s_{m_k}(-\eta) | \\ 
    = (-1)^{ \frac12 + \frac32 +- \dots + (k - \frac12) } \sum
      | s_{l_1}(\epsilon) , s_{l_2}(\epsilon) , \dots , 
        s_{l_{\Omega}}(\epsilon) |
      | c_{m_1}(\eta) , c_{m_2}(\eta) , \dots ,
        c_{m_k}(\eta) | .
\end{multline}
By~\cite{ref:Wey39,ref:Lit40}
\begin{equation}\label{eq:chi-o-d}
  \chi_{[\lambda]} = \frac
    { | s_{l_1}(\epsilon) , s_{l_2}(\epsilon) , \dots , 
        s_{l_{\Omega}}(\epsilon) | }
    { | s_{\frac12}(\epsilon) , s_{\frac32}(\epsilon) , \dots ,
      s_{\Omega-\frac12}(\epsilon) | } , \quad
  \chi_{[\mu]} = \frac
    { | c_{m_1}(\eta) , c_{m_2}(\eta) , \dots , 
        c_{m_k}(\eta) | }
    { | c_0(\eta) , c_1(\eta) , \dots 
      c_{k-1}(\eta) | } ,
\end{equation}
the identity~\eqref{eq:charsum} follows.

\section{\label{sec:spgl}Symplectic and general linear Lie algebras}

\subsection{\label{sec:sp}Symplectic Lie algebras}

The algebras $\lsp(d)$ and $\lsp(2 k)$ are very similar to $\lo(d)$
and $\lo(2 k)$. A Lie algebra of generators $x$ of symplectic
transformations of the 1-kind configuration space is defined by the
conservation of a non-singular antisymmetric tensor
$\langle p q | g \rangle = - \langle q p | g \rangle$ in the sense of
Eq.~\eqref{eq:cons_s}. The non-singularity of $g$ requires that $d$ is
even. In terms of these $x$ the members $X$ of $\lsp(d)$ are then
defined by Eq.~\eqref{eq:S}. The index $p$ of basic, orthonormal
1-kind states $|p \rangle$ may take the values
$-\Omega, -\Omega + 1, \dots, \Omega$ with 0 omitted. It will be
assumed that these states can be so chosen that
\begin{equation}\label{eq:g_diag-a}
  \langle p q | g \rangle = \sigma_p \, \delta_{p,-q} 
\end{equation}
with $\sigma_p = \pm 1$ for $p \gtrless 0$. In atomic and nuclear
physics, the 1-kind configuration space often carries half-integral
angular momentum quantum numbers $j$, and $g$ is chosen to be the
Wigner metric
\begin{equation}~\label{eq:Wig-j}
  \langle \alpha j m , \alpha' j' m'| g \rangle
    = (-)^{j+m} \delta_{\alpha\alpha'} \delta_{jj'} \delta_{m,-m'}
\end{equation}
in a notation as in Eq.~\eqref{eq:Wig-l}. This $g$ can be given the
form~\eqref{eq:g_diag-a} by a unitary transformation of the 1-kind
configuration space.

The operators $x_{pq}$ defined by
\begin{equation}
  x_{pq} | - q \rangle = \sigma_p | p \rangle, \quad
  x_{pq} | - p \rangle = \sigma_q | q \rangle, \quad p \ne q , \quad
  x_{pp} | - p \rangle = 2 \sigma_p | p \rangle, \quad
  x_{pq} | r \rangle = 0, \quad r \ne -p , -q ,
\end{equation}
satisfy $x_{pq} = x_{qp}$. Those with $p \ge q$ form a basis for the
Lie algebra. A basis for a Cartan subalgebra which gives the roots of
Ref.~\onlinecite{ref:Rac51} is formed by the operators $x_{p,-p}$ with
$p > 0$. Renormalizing the Killing form $(x,x')$ so that these
operators have squared Killing norm 1 results in
\begin{equation}
  ( x_{pq} , x_{-p,-q} ) = - \sigma_p \sigma_q , \quad p \ne q , \quad
  ( x_{pp} , x_{-p,-p} ) = - 2 , \quad
  \text{otherwise } ( x_{pq} , x_{rs} ) = 0 .
\end{equation}
This gives the Casimir operator
\begin{equation}\label{eq:Co-d-sp}
  C_\text{$\lsp(d)$}
    = - \tfrac12 \sum_{pq} \sigma_p \sigma_q X_{pq}  X_{-q,-p} .
\end{equation}

An analysis like that of Sec.~\ref{sec:ncons} gives the maximal Lie
algebra spanned by commutators $[\phi,\chi]$ that commute with every
$X$. Its elements of the form~\eqref{eq:dag-none} are given by
Eq.~\eqref{eq:Y+-} while those of the form~\eqref{eq:dag-dag} are
given by Eq.~\eqref{eq:dd-bas-g}, which for the
metric~\eqref{eq:g_diag-a} evaluates to
\begin{equation}\label{eq:dd-bas-sp}
  Y_{\tau\upsilon}
    = \sum_p \sigma_p a^\dagger_{p\tau} a^\dagger_{-p,\upsilon} .
\end{equation}
Including the Hermitian conjugates $Y_{-\tau,-\upsilon}$ of the latter
renders the set maximal. With indices $\alpha,\beta,\dots$ as in
Eq.~\eqref{eq:com}, one may verify the commutation relations
\begin{equation}\label{eq:com-sp}
  [ Y_{\alpha\beta} , Y_{\gamma\delta} ]
    = \sigma_\gamma \delta_{\beta,-\gamma} Y_{\alpha\delta}
      + \sigma_\delta \delta_{\beta,-\delta} Y_{\alpha\gamma}
      + \sigma_\gamma \delta_{\alpha,-\gamma} Y_{\beta\delta}
      + \sigma_\delta \delta_{\alpha,-\delta} Y_{\beta\gamma} ,
\end{equation}
which identify the resulting Lie algebra as an $\lsp(2 k)$
algebra.~\cite{ref:Rac51} The operators $Y_{\tau,-\tau}$ again form a
basis for a Cartan subalgebra which gives the roots of
Ref.~\onlinecite{ref:Rac51}, and renormalizing the Killing form so
that these operators have squared Killing norm 1 results in an
expression for the Casimir operator analogous to
Eq.~\eqref{eq:Co-d-sp}. The discussion in Sec.~\ref{sec:thrm} applies
almost verbatim to the $\lsp(d)$-$\lsp(2 k)$ duality, the only
differences being that only integral parts occur, every partition
describes an irreducible representation, and Eqs.~\eqref{eq:l-e} and
\eqref{eq:l-o} are replaced by
\begin{equation}
  l_p = \lambda_p + p ,
\end{equation}
and the sequences~\eqref{eq:seq-e} and \eqref{eq:seq-o} by
\begin{equation}
  1, 2, ... \, , \Omega + k .
\end{equation}

\subsection{\label{sec:gl}General linear Lie algebras}

The general linear Lie algebra $\lgl(d)$ is induced in the manner of
Secs.~\ref{sec:cons} and \ref{sec:sp} by the Lie algebra of arbitrary
linear transformations $x$ of the 1-kind configuration space. The
latter has a basis of operators $x_{pq}$ defined by
\begin{equation}
  x_{pq} | q \rangle = | p \rangle, \quad
  x_{pq} | r \rangle = 0, \quad r \ne q .
\end{equation}
This Lie algebra is not semi-simple, but its special linear subalgebra
of traceless linear transformations is simple. The Killing form of the
latter can be renormalized to be the restriction to the subalgebra of
the form
\begin{equation}\label{eq:Kil-gl}
  ( x_{pq} , x_{rs} ) = \delta_{pq,sr}
\end{equation}
defined on the general linear Lie algebra. It is therefore customary
in nuclear theory to define a Killing form for the general linear Lie
algebra by Eq.~\eqref{eq:Kil-gl}, and accordingly a Casimir operator
$C_\text{$\lgl(d)$}$ of $\lgl(d)$ by~\cite{ref:Edm52}
\begin{equation}\label{eq:Co-d-gl}
  C_\text{$\lgl(d)$} = \sum_{pq}  X_{pq}  X_{qp} .
\end{equation}
Because its difference from the Casimir operator $C_\text{$\lsl(d)$}$
of the special linear subalgebra $\lsl(d)$ is a term in an operator
which spans the commuting 1-dimensional extension from $\lsl(d)$ to
$\lgl(d)$, the operator $C_\text{$\lgl(d)$}$ is a $\lgl(d)$ invariant.

The irreps of $\lgl(d)$ that are realized on $\Phi$ are described by
partitions with at most $d$ non-negative, integral parts. Unlike the
cases of the $\lo(d)$ and $\lsp(d)$, I obey in this case the
convention, which is to number the parts in non-increasing order,
\begin{equation}
  \lambda_1 \ge \lambda_2 \ge \dots 
    \ge \lambda_d \ge 0
\end{equation}

The Lie algebra $\lgl(k)$ results from interchanging the roles of the
indices $p$ and $\tau$. Specifically, Eq.~\eqref{eq:S} is replaced by
\begin{equation}\label{eq:Y-gl}
  Y = \sum_{\tau\upsilon p} \langle \tau | y | \upsilon \rangle
    a^\dagger_{p\tau} a_{p\upsilon} .
\end{equation}
where $\langle \tau | y | \upsilon \rangle$ is any $k \times k$
matrix. This renders $\lgl(k)$ analogous to $\lgl(d)$ in every
respect. The theorem of $\lgl$-$\lgl$ duality was stated already in
the introduction. The connected irreps have conjugate Young diagrams.
Therefore the $\lgl(d)$ parts do not exceed $k$ and the $\lgl(k)$
parts do not exceed $d$.

\section{\label{sec:cas}Relations of Casimir invariants}

\subsection{\label{sec:cas-int}Derivations from duality}

For any one of the Lie algebras $\lgr(d)$ introduced above, the
eigenvalue of the Casimir operator $C_{\text{$\lgr(d)$}}$ on the
carrier space of an irrep $[\lambda]$ is denoted by
$C_{\text{$\lgr(d)$}}([\lambda])$ and referred to as the Casimir
invariant of that irrep. For the semi-simple Lie algebras $\lo(d)$ and
$\lsp(d)$, a general formula derived by
Racah~\cite{ref:Rac50,ref:Rac51} gives
\begin{equation}\label{eq:Rac}
  C_{\text{$\lgr(d)$}}([\lambda])
    = \sum_p \lambda_p ( 2 l_p - \lambda_p ) .
\end{equation}
For $\lo(d)$ and even $d$, I have here temporarily allowed either sign
of $\lambda_1$ when it differs from 0. However, in this case the
expression~\eqref{eq:Rac} does not depend on the sign of $\lambda_1$,
so the Casimir invariant is common to both irreducible constituents of
the representation described by the partition $[\lambda]$ defined in
cf. Sec~\ref{sec:rep}, which makes it, actually, a function of this
partition.

The formula~\eqref{eq:Rac} can be given a geometric form: Assume the
usual arrangement of the $[\lambda]$ Young diagram with horizontal
rows of lengths $\lambda_p$, left justified from bottom to top in the
order of $p$, and assume that the cells are unit squares except for
the half-width cells of the spin representations of $\lo(d)$. Then
place the origin of a Cartesian coordinate system at the distance
$d/2$ vertically below the upper left corner of the diagram. It is
straightforward to show that Eq.~\eqref{eq:Rac} can then be written
\begin{equation}\label{eq:int}
  C_{\text{$\lgr(d)$}}([\lambda]) = \int ( 2 ( x + y ) \mp 1 ) dx dy
\end{equation}
with the integration extended over the area of the diagram. The upper
and lower signs apply to $\lo(d)$ and $\lsp(d)$, respectively.

The configuration where the $[\lambda]$ and $[\mu]$ diagrams fill a
rectangle of width $k$ and height $d/2$ as described in
Sec.~\ref{sec:stmnt} results from first placing them so that their
coordinate systems coincide and then reflecting the $[\mu]$ diagram in
the line $x = y$. As the integral~\eqref{eq:int} is invariant under
this reflection, one gets $C_{\text{$\lgr(d)$}}([\lambda]) +
C_{\text{$\lgr(2 k)$}}([\mu])$ by integration over the entire
rectangle. Hence, when $[\lambda]$ and $[\mu]$ are dual
representations,
\begin{equation}\label{eq:cassum}
  C_{\text{$\lgr(d)$}}([\lambda]) + C_{\text{$\lgr(2 k)$}}([\mu])
  = k^2 \frac d 2 + k \left( \frac d 2 \right)^{\!2} \!\mp k \frac d 2
  = \tfrac14 k d ( d + 2 k \mp  2 ) .
\end{equation}

The principle of derivation of Racah's formula can be extended to
$\lgl(d)$ with the result~\cite{ref:Edm52}
\begin{equation}\label{eq:Rac-gl}
  C_{\text{$\lgl(d)$}}([\lambda])
    = \sum_p \lambda_p ( \lambda_p + d + 1 - 2 p ) .
\end{equation}
When the origin of the coordinate system is now placed at the upper
left corner of the Young diagram and the $y$ axis is turned downwards,
this becomes
\begin{equation}\label{eq:int-gl}
  C_{\text{$\lgl(d)$}}([\lambda]) = \int ( 2 ( x - y ) + d ) dx dy
\end{equation}
Using that conjugate Young diagrams are reflections of one another and
the area of the diagram equals $n$, one
gets for dual representations $[\lambda]$ and $[\mu]$ that
\begin{equation}\label{eq:cassum-gl}
  C_{\text{$\lgl(d)$}}([\lambda]) + C_{\text{$\lgl(k)$}}([\mu])
    = ( d + k ) n .
\end{equation}

Before closing this section, I notice that the integral
formulas~\eqref{eq:int} and \eqref{eq:int-gl} allow simple derivations
of formulas for $C_{\text{$\lgr(d)$}}([\lambda])$ in terms of the
column heights $\nu_i$ of the Young diagram, numbered by $i$ from the
left, not counting the half-width column in the case of spin
representations of $\lo(d)$. One gets
\begin{equation}\begin{gathered}
  C_{\text{$\lgr(d)$}}([\lambda]) = \sum_i
    \nu_i ( d - 1 \mp 1 + 2 i - \nu_i ) \qquad
  \text{for $\lsp(d)$ and non-spin representations of $\lo(d)$} , \\
  C_{\text{$\lo(d)$}}([\lambda]) = \sum_i
    \nu_i ( d - 1 + 2 i - \nu_i)
     + \tfrac12 \Omega ( d - \Omega  - \tfrac12 ) \qquad
  \text{for spin representations} , \\
  C_{\text{$\lgl(d)$}}([\lambda]) = \sum_i
    \nu_i ( d - 1 + 2 i - \nu_i ) .
\end{gathered}\end{equation}

\subsection{\label{sec:alg}Algebraic derivation}

For $k = 4$, Eqs.~\eqref{eq:cassum} and \eqref{eq:cassum-gl} give
\begin{equation}
  C_{\text{$\lo(d)$}}([\lambda]) + C_{\text{$\lo(8)$}}([\mu])
    =  d ( d + 6 ) , \quad
  C_{\text{$\lgl(d)$}}([\lambda]) + C_{\text{$\lgl(4)$}}([\mu])
    = ( d + 4 ) n .
\end{equation}
These relations were derived previously by Kota and Castilho
Alcar\'as~\cite{ref:Kot06} using an algebraic method. It is
instructive to compare their method with that of the preceding
subsection. I confine myself to deriving Eq.~\eqref{eq:cassum} for the
case of the $\lo$-$\lo$ duality in this way. One aim of showing this
derivation is to demonstrate that such algebraic work gets simpler
without recourse to the Wigner metric~\eqref{eq:Wig-l} and angular
momentum algebra.

I use the expressions~\eqref{eq:Co-d-o} and \eqref{eq:Co-2k} for the
Casimir operators. Substitution by Eq.~\eqref{eq:S} gives
\begin{multline}\label{eq:COdr}
  C_\text{O$(d)$} = \tfrac12 \sum_{pq\tau\upsilon}
    ( a_{p\tau}^\dagger a_{q\tau} - a_{q\tau}^\dagger a_{p\tau} )
    ( a_{q\upsilon}^\dagger a_{p\upsilon}
      - a_{p\upsilon}^\dagger a_{q\upsilon} )
    = \sum_{pq\tau\upsilon} a_{p\tau}^\dagger a_{q\tau}
      ( a_{q\upsilon}^\dagger a_{p\upsilon}
        - a_{p\upsilon}^\dagger a_{q\upsilon} ) \\
    = \sum_{pq\tau\upsilon} a_{p\tau}^\dagger
      ( a_{q\upsilon}^\dagger a_{p\upsilon}
        - a_{p\upsilon}^\dagger a_{q\upsilon} ) a_{q\tau}
      + \sum_{pq\tau} a_{p\tau}^\dagger a_{p\tau}
      - \sum_{p\tau} a_{p\tau}^\dagger a_{p\tau}
   = \sum_{pq\tau\upsilon} a_{p\tau}^\dagger
      ( a_{q\upsilon}^\dagger a_{p\upsilon}
        - a_{p\upsilon}^\dagger a_{q\upsilon} ) a_{q\tau}
      + d n - n .
\end{multline}
By the commutations relations~\eqref{eq:com} and Eqs.~\eqref{eq:Y+-}
and \eqref{eq:dd-bas}, one gets
\begin{multline}\label{eq:CO2kr}
 C_\text{O$(2 k)$} = \sum_{\tau\upsilon}
   ( \tfrac12 ( Y_{\tau\upsilon} Y_{-\upsilon,-\tau}
      + Y_{-\upsilon,-\tau} Y_{\tau\upsilon} )
      + Y_{\tau,-\upsilon} Y_{\upsilon,-\tau} ) \\
    = \sum_{\tau\upsilon} ( Y_{\tau\upsilon} Y_{-\upsilon,-\tau}
      - \tfrac12 ( 1 - \delta_{\tau\upsilon} )
        ( Y_{\tau,-\tau} + Y_{\upsilon,-\upsilon} )
      + Y_{\tau,-\upsilon} Y_{\upsilon,-\tau} )
    = \sum_{\tau\upsilon} ( Y_{\tau\upsilon} Y_{-\upsilon,-\tau}
      + Y_{\tau,-\upsilon} Y_{\upsilon,-\tau} )
      - (k - 1) \sum_\tau  Y_{\tau,-\tau} \\
    = \sum_{pq\tau\upsilon} ( a_{p\tau}^\dagger a_{p\upsilon}^\dagger
        a_{q\upsilon} a_{q\tau}
      + ( a_{p\tau}^\dagger a_{p\upsilon}
          - \tfrac12 \delta_{\tau\upsilon} )
        ( a_{q\upsilon}^\dagger a_{q\tau}
          - \tfrac12 \delta_{\tau\upsilon} ) )
      - (k - 1) \sum_{p\tau} ( a_{p\tau}^\dagger a_{p\tau}
        - \tfrac12 ) \\
    = \sum_{pq\tau\upsilon} a_{p\tau}^\dagger
        ( a_{p\upsilon}^\dagger a_{q\upsilon}
          - a_{q\upsilon}^\dagger a_{p\upsilon} ) a_{q\tau}
      + \sum_{p\upsilon\tau} a_{p\tau}^\dagger a_{p\tau}
      - \sum_{pq\tau} a_{p\tau}^\dagger a_{p\tau}
      + \tfrac14 d^2 k
      - (k - 1) \sum_{p\tau} ( a_{p\tau}^\dagger a_{p\tau}
        - \tfrac12 ) \\
    = \sum_{pq\tau\upsilon} a_{p\tau}^\dagger
      ( a_{p\upsilon}^\dagger a_{q\upsilon} 
        - a_{q\upsilon}^\dagger a_{p\upsilon} ) a_{q\tau}
      + k n - d n \\ + \tfrac14 d^2 k
      - (k - 1) ( n - \tfrac12 d k)
    = \sum_{pq\tau\upsilon} a_{p\tau}^\dagger
      ( a_{p\upsilon}^\dagger a_{q\upsilon}
        - a_{q\upsilon}^\dagger a_{p\upsilon} ) a_{q\tau}
        + n - d n + \tfrac14 k d ( d + 2 k - 2 ) .
\end{multline}
Adding Eqs.~\eqref{eq:COdr} and \eqref{eq:CO2kr} results in
Eq.~\eqref{eq:cassum} once again.

For the $\lsp$-$\lsp$ duality, Eq.~\eqref{eq:cassum} was derived in
this way in Ref.~\onlinecite{ref:Hel61}, and Eq.~\eqref{eq:cassum-gl}
results after some commutations when the term
$\sum_{pq\tau\upsilon} a_{p\tau}^\dagger a_{p\upsilon}^\dagger
a_{q\upsilon} a_{q\tau}$ is moved from the sum in Eq.~\eqref{eq:CO2kr}
to the sum in Eq.~\eqref{eq:COdr}. (When the
metric~\eqref{eq:g_diag-s} is equivalent to the Wigner
metric~\eqref{eq:Wig-l}, this term is a pairing interaction, and
analogously in the symplectic case. Considering the sums with and
without it then also gives expressions for the eigenvalue of
this interaction in terms of Casimir
invariants.~\cite{ref:Rac49,ref:Edm52,ref:Rac52,ref:Hel61,ref:Flo64a,
ref:Flo64b,ref:Kot06,ref:Row07})

\section{\label{sec:phys}Orthogonal-orthogonal dualities in physics}

The only case of an $\lo$-$\lo$ duality which is known to me to have
been discussed in relation to actual physical systems is the one where
$\lo(2 k)$ is the $\lo(8)$ algebra of Flowers and Szpikowski. It seems 
reasonable to address the question whether other cases than $k = 4$
could have meaningful applications.

For $k = 1$ the 1-kind configuration space contains all possible
states of one fermion. One may think of an atomic or nuclear shell.
The $\lo(d)$ representations have Young diagrams with at most 1
column, so their carrier spaces are characterized by values
$n \le d/2$ and is isomorphic to the space of all antisymmetric
functions of $p_1, \dots, p_n$. The carrier spaces of the $\lo(2 k)$
representations have two basic states with $n = d/2 \pm \mu_1$ except
when $\mu_1 = 0$. The $\lo$-$\lo$ duality thus imposes identical
structures on the two subspaces of $\Phi$ with these numbers of
fermions. This reflects the well known particle-hole
symmetry,~\cite{ref:Con35,ref:Rac42,ref:Bel59} which thus turns out to
be encoded in the $\lo$-$\lo$ duality. For $\mu_1 = 0$, corresponding
to $n = d/2$, particle-hole conjugation maps the $\lo(d)$
representation space onto itself, and the two irreps into which it
splits in this case are characterized by opposite particle-hole
conjugation parities.

For $k = 2$ the states $|p\rangle$ could be states of motion in space,
and $\tau = 1$ and 2 could denote spin directions $\downarrow$ and
$\uparrow$. This could describe an atomic shell or a spin saturated
shell of neutron orbits or of proton orbits in a semi-magic nucleus.
It is well known that $\lo(2 k) = \lo(4)$ is not simple but splits
into two commuting $\lo(3)$ algebras.~\cite{ref:Rac51,ref:Jac62} These
can be chosen to be spanned by the components of the total spin
\begin{equation}\label{eq:S-O4}
  {\bm S} = \sum_{p\tau \upsilon}
    \langle \tau | {\bm s} | \upsilon \rangle
    a^\dagger_{p\tau} a_{p\upsilon} ,
\end{equation}
where $\bm s$ is the spin vector acting on the span of
$|\!\!\downarrow\rangle$ and $|\!\!\uparrow\rangle$, and by the
components
\begin{equation}\label{eq:Q-O4}
  Q_0 = \tfrac12 ( n - d ), \quad Q_- = \sum_p a_{p\downarrow}
  a_{p\uparrow} , \quad Q_+ = Q_-^\dagger ,
\end{equation}
of a ``spin quasi-spin'' $\bm Q$. I denote accordingly the two
$\lo(3)$ algebras by $\lo(3)_S$ and $\lo(3)_Q$. An irrep of the
$\lo(4)$ algebra is described by a pair of the single parts $S$ and $Q$
of the $\lo(3)$ partitions (related to the Casimir invariants
${\bm S}^2 = S(S+1)$ and ${\bm Q}^2 = Q(Q+1)$).

The Lie algebra of operators $Y$ is isomorphic to that of
transformations
\begin{equation}
  \cal Y: \phi \mapsto [ Y , \phi ]
\end{equation}
of the space of fermion fields $\phi$. As the former faithfully
represents the $d = 1$ Lie algebras, this then holds for the latter,
as well. For $d = 1$ and $k = 2$ the space of fermion fields $\phi$ is
4-dimensional. Swapping $a_{1\downarrow}$ and
$a^\dagger_{1\downarrow}$ in the composition of $\phi$ is a reflection
of this space. In combination with the transformations $\mathcal Y$,
which generate the proper orthogonal group $SO(4)$, it therefore
generates the full orthogonal group $O(4)$. Swapping $a_{1\downarrow}$
and $a^\dagger_{1\downarrow}$ in the composition of $\phi$ is
equivalent to doing so in the composition of the operator $Y$, which
maps to swapping $a_{p\downarrow}$ and $a^\dagger_{p\downarrow}$ in
the composition of $Y$ for every $p$ when $d$ is arbitrary. The latter
operation is equivalent, in turn, to the operation $P$ of swapping
emptiness and occupation of the states $| p \!\! \downarrow \rangle$
in the composition of a state in $\Phi$. Appending $P$ to $\lo(4)$
therefore generates a representation of $O(4)$ on $\Phi$.

Swapping $a_{p\downarrow}$ and $a^\dagger_{p\downarrow}$ in the
expressions~\eqref{eq:S-O4} and \eqref{eq:Q-O4} swaps $\bm S$ and
$\bm Q$. It follows that the irreps of $O(4)$ are direct sums of two
$\lo(3)_S \oplus \lo(3)_Q$ irreps with swapped, different $S$ and $Q$
or a single such irrep with equal $S$ and $Q$. The weight components
$w_\tau$ are the eigenvalues of $Y_{\tau,-\tau}$ in the representation
module, which are given, by Eqs.~\eqref{eq:Y+-}, \eqref{eq:S-O4} and
\eqref{eq:Q-O4}, by $w_\tau = Q_0 \mp S_0$ for $\tau = 1$ and~2. The
highest weight of an $\lo(3)_S \oplus \lo(3)_Q$ irrep then has
$w_\tau = Q \mp S$, so $(\mu_1,\mu_2) = (|S - Q|,S + Q)$. The quantum
numbers $S$ and $Q$ thus determine the partition $[\mu]$ and, in turn,
by the $\lo$-$\lo$ duality, the partition $[\lambda]$. Basic states
with a given $n$ belong to representations of $\lo(d) \oplus \lo(3)_S$
described by $[\lambda]$ and $S$, and $\lo(3)_Q$ connects analogous
states of this form with $d - 2 Q \le n \le d + 2 Q$ and even
$n - (d - 2 Q)$. In this way the spin quasi-spin $\bm Q$ is analogous
to Kerman's quasispin pertaining to spin non-saturated shells. Like
Flowers's reduced isospin, the quantum number
$\sigma = (\mu_2 - \mu_1)/2 = \min (S,Q)$ could be called a reduced
spin.

The seniority $v_\text R$ defined in Ref.~\onlinecite{ref:Rac43} is
the first $n$ in a chain of connected states, so
$v_\text R = d - 2 Q$. It equals the area $v = d - 2 \max(S,Q)$ of the
$\lo(d)$ Young diagram only when $S \le Q$. (It is the height of the
1-column Young diagram of the enclosing $\lsp(2 d)$ irrep dual to the
$\lo(3)_Q$ irrep.~\cite{ref:Rac49}) When $S \le Q$ one has
$v_\text R,S = v,\sigma$, so $C_{\text{$\lo(d)$}}([\lambda])$ is the
same function of $v_\text R$ and $S$ as of $v$ and $\sigma$. Because
$C_{\text{$\lo(4)$}}([\mu])$, and therefore
$C_{\text{$\lo(d)$}}([\lambda])$ by Eq.~\eqref{eq:cassum}, is a
polynomial in $S$ and $Q$, this then holds also for $S > Q$. As a
result, when $v_\text R$ and $S$ are taken as the independent
variables, a term $-2S(S+1)$ cancels out in the difference
$C_{\text{$\lgl(d)$}}([\lambda']) - C_{\text{$\lo(d)$}}([\lambda])$,
where $[\lambda']$ is the enclosing $\lgl(d)$ irrep. This explains
that the eigenvalue of the pairing interaction considered in
Ref.~\onlinecite{ref:Rac43} depends only on $n$ and $v_\text R$ and
not on $S$ and $\sigma$, cf. Eq.~(50) there. When $v$ rather than
$v_\text R$ is taken as the independent valiable, this eigenvalue
depends on $S$ and $\sigma$ in a way analogous to the dependence of
such an eigenvalue on isospin and reduced isospin in the system of
neutrons and protons sharing a configuration space.\cite{ref:Edm52}

The algebra $\lo(3)_S$ is identical to the algebra $\lsl(2)$ of
operators $Y$ in Eq.~\eqref{eq:Y-gl} with a traceless $y$. For $k > 2$
one may similarly decompose the $\lo(2 k)$ representation into
$\lsl(k)$ irreps, which may be combined with the $\lo(d)$
representation to classify basic states with a fixed $n$. The
$\lo(2 k)$ algebra then connects such representation spaces with
different $n$. For $k > 2$ the $\lo(2 k)$ algebra is simple, so the
$\lsl(k)$ algebra has no commuting subalgebra like $\lo(3)_Q$.
Therefore, while the connected representations have equal $\lo(d)$
factors, their $\lsl(k)$ factors may differ.

As an example, the $\lsl(4)$ subalgebra of the $\lo(8)$ algebra of
Flowers and Szpikowski was identified by Wigner in the early days of
nuclear theory.~\cite{ref:Wig37} Traditionally, it is referred to by
nuclear physicists as $SU(4)$, and its irreps are decomposed further
into irreps of a direct sum $\lsl(2)_S \oplus \lsl(2)_T$ of Lie
algebras of spin and isospin. Kota and Castilho Alcar\'as determined
the branching of $\lo(8)$ representations into $\lsl(4)$, or
$\lgl(4)$, irreps by analyzing the dual branching of $\lgl(d)$ irreps
into $\lo(d)$ representations.~\cite{ref:Kot06}

The discussion above of systems of fermions with spin $s = 1/2$
may be generalized to fermions with $s > 1/2$. One
then has $k = 2 s + 1 > 2$. Because fermion spins are half-integral,
$k$ is even. In principle a symmetric metric $g$ may be defined on any
1-particle configuration space, including configuration spaces of
single fermions. But in the fermion case, unlike the Wigner
metric~\eqref{eq:Wig-l}, $g$ will not be rotationally invariant. The
Wigner metric~\eqref{eq:Wig-j} on a space of 1-fermion states is
antisymmetric and so gives rise to a symplectic Lie algebra. This lack
of rotational invariance seems to bar the relevance of the $\lo$-$\lo$
duality to systems with an odd number of fermion kinds (more than one)
such as the three colors or the three low-mass flavors of quarks.

\section{\label{sec:sum}Summary}

The Fock space $\Phi$ spanned by all possible states of any number of
fermions of $k$ different kinds inhabiting a common configuration
space of dimension $d$ is considered. The theorem of
orthogonal-orthogonal duality states that $\Phi$ is composed of outer
products $\text{X}_{[\lambda]} \otimes \Psi_{[\mu]}$, where
$\text{X}_{[\lambda]}$ carries the representation of $\lo(d)$
associated with the partition $[\lambda]$, and $\Psi_{[\mu]}$
similarly, involving the Lie algebra $\lo(2 k)$. Here $\lo(d)$ denotes
the Lie algebra of infinitesimal orthogonal transformations of a
vector space of dimension $d$. Each pair of partitions $[\lambda]$ and
$[\mu]$ which obey a certain criterion appears exactly once. This
criterion can be expressed in a simple geometric form: The $[\lambda]$
Young diagram and a reflected copy of the $[\mu]$ Young diagram fill a
rectangle of width $k$ and height $d/2$ without overlap. The algebra
$\lo(d)$ conserves the number of fermions in the system and the
algebra $\lo(2 k)$ does not.

An analogous symplectic-symplectic duality was proven by Helmers
almost 60 years ago, but only recently was the orthogonal-orthogonal
duality proven by a very different method.~\cite{ref:Row11} I have
presented a proof by the same method as applied by Helmers, using the
known expressions for the characters of orthogonal Lie algebras.

Besides the orthogonal-orthogonal and symplectic-symplectic dualities,
the authors of Ref.~\onlinecite{ref:Row11} list and prove a
unitary-unitary duality. This expresses the following fact, known
since long ago: The space of antisymmetric functions of $n$ pairs of
variables $(p,\tau)$, where $p$ takes $d$ values and $\tau$ takes $k$
values, is composed of antisymmetrized products of a function $\psi$
of the variables $p$ and a function $\chi$ of the variables $\tau$. In
this direct sum, each antisymmetrized product of an irrep of the
general linear Lie algebra $\lgl(d)$ (or its unitary subalgebra
$\lu(d)$) acting on $\psi$, and an irrep of the general linear Lie
algebra $\lgl(k)$ (or its unitary subalgebra $\lu(k)$) acting on
$\chi$, appear exactly once according to the criterion that the two
irreps must be associated with Young diagrams that are reflections of
one another.

In the present article, the geometric relation between the Young
diagrams involved in each of these three duality theorems was shown to
allow very simple derivations of linear relations between the Casimir
invariants of the connected representations. Only particular cases of
these linear relations seem to have been presented previously in the
literature.

A final section addressed the question of possible applications the
orthogonal-orthogonal duality to the analysis of actual physical
systems other than such already considered in the literature.

\bibliography{dual}

\begin{thebibliography}{41}%
\makeatletter
\providecommand \@ifxundefined [1]{%
 \@ifx{#1\undefined}
}%
\providecommand \@ifnum [1]{%
 \ifnum #1\expandafter \@firstoftwo
 \else \expandafter \@secondoftwo
 \fi
}%
\providecommand \@ifx [1]{%
 \ifx #1\expandafter \@firstoftwo
 \else \expandafter \@secondoftwo
 \fi
}%
\providecommand \natexlab [1]{#1}%
\providecommand \enquote  [1]{``#1''}%
\providecommand \bibnamefont  [1]{#1}%
\providecommand \bibfnamefont [1]{#1}%
\providecommand \citenamefont [1]{#1}%
\providecommand \href@noop [0]{\@secondoftwo}%
\providecommand \href [0]{\begingroup \@sanitize@url \@href}%
\providecommand \@href[1]{\@@startlink{#1}\@@href}%
\providecommand \@@href[1]{\endgroup#1\@@endlink}%
\providecommand \@sanitize@url [0]{\catcode `\\12\catcode `\$12\catcode
  `\&12\catcode `\#12\catcode `\^12\catcode `\_12\catcode `\%12\relax}%
\providecommand \@@startlink[1]{}%
\providecommand \@@endlink[0]{}%
\providecommand \url  [0]{\begingroup\@sanitize@url \@url }%
\providecommand \@url [1]{\endgroup\@href {#1}{\urlprefix }}%
\providecommand \urlprefix  [0]{URL }%
\providecommand \Eprint [0]{\href }%
\providecommand \doibase [0]{http://dx.doi.org/}%
\providecommand \selectlanguage [0]{\@gobble}%
\providecommand \bibinfo  [0]{\@secondoftwo}%
\providecommand \bibfield  [0]{\@secondoftwo}%
\providecommand \translation [1]{[#1]}%
\providecommand \BibitemOpen [0]{}%
\providecommand \bibitemStop [0]{}%
\providecommand \bibitemNoStop [0]{.\EOS\space}%
\providecommand \EOS [0]{\spacefactor3000\relax}%
\providecommand \BibitemShut  [1]{\csname bibitem#1\endcsname}%
\let\auto@bib@innerbib\@empty
\bibitem [{\citenamefont {Helmers}(1961)}]{ref:Hel61}%
  \BibitemOpen
  \bibfield  {author} {\bibinfo {author} {\bibfnamefont {K.}~\bibnamefont
  {Helmers}},\ }\href@noop {} {\bibfield  {journal} {\bibinfo  {journal} {Nucl.
  Phys.}\ }\textbf {\bibinfo {volume} {23}},\ \bibinfo {pages} {594} (\bibinfo
  {year} {1961})}\BibitemShut {NoStop}%
\bibitem [{\citenamefont {Bohr}, \citenamefont {Mottelson},\ and\ \citenamefont
  {Pines}(1958)}]{ref:Boh58}%
  \BibitemOpen
  \bibfield  {author} {\bibinfo {author} {\bibfnamefont {A.}~\bibnamefont
  {Bohr}}, \bibinfo {author} {\bibfnamefont {B.~R.}\ \bibnamefont {Mottelson}},
  \ and\ \bibinfo {author} {\bibfnamefont {D.}~\bibnamefont {Pines}},\
  }\href@noop {} {\bibfield  {journal} {\bibinfo  {journal} {Phys. Rev.}\
  }\textbf {\bibinfo {volume} {110}},\ \bibinfo {pages} {936} (\bibinfo {year}
  {1958})}\BibitemShut {NoStop}%
\bibitem [{\citenamefont {Bogolyubov}(1958)}]{ref:Bog58}%
  \BibitemOpen
  \bibfield  {author} {\bibinfo {author} {\bibfnamefont {N.~N.}\ \bibnamefont
  {Bogolyubov}},\ }\href@noop {} {\bibfield  {journal} {\bibinfo  {journal}
  {Dokl. Akad. Nauk SSSR}\ }\textbf {\bibinfo {volume} {119}},\ \bibinfo
  {pages} {52} (\bibinfo {year} {1958})},\ \bibinfo {note} {[Sov. Phys. Dokl.
  \textbf{3}, 279 (1958)]}\BibitemShut {NoStop}%
\bibitem [{\citenamefont {Solov'yov}(1958{\natexlab{a}})}]{ref:Sol58a}%
  \BibitemOpen
  \bibfield  {author} {\bibinfo {author} {\bibfnamefont {V.~G.}\ \bibnamefont
  {Solov'yov}},\ }\href@noop {} {\bibfield  {journal} {\bibinfo  {journal}
  {Dokl. Akad. Nauk SSSR}\ }\textbf {\bibinfo {volume} {123}},\ \bibinfo
  {pages} {437} (\bibinfo {year} {1958}{\natexlab{a}})},\ \bibinfo {note}
  {[Sov. Phys. Dokl. \textbf{3}, 1176 (1958)]}\BibitemShut {NoStop}%
\bibitem [{\citenamefont {Solov'yov}(1958{\natexlab{b}})}]{ref:Sol58b}%
  \BibitemOpen
  \bibfield  {author} {\bibinfo {author} {\bibfnamefont {V.~G.}\ \bibnamefont
  {Solov'yov}},\ }\href@noop {} {\bibfield  {journal} {\bibinfo  {journal}
  {Dokl. Akad. Nauk SSSR}\ }\textbf {\bibinfo {volume} {123}},\ \bibinfo
  {pages} {652} (\bibinfo {year} {1958}{\natexlab{b}})},\ \bibinfo {note}
  {[Sov. Phys. Dokl. \textbf{3}, 1197 (1958)]}\BibitemShut {NoStop}%
\bibitem [{\citenamefont {Bardeen}, \citenamefont {Cooper},\ and\ \citenamefont
  {Schrieffer}(1957{\natexlab{a}})}]{ref:Bar57a}%
  \BibitemOpen
  \bibfield  {author} {\bibinfo {author} {\bibfnamefont {J.}~\bibnamefont
  {Bardeen}}, \bibinfo {author} {\bibfnamefont {L.~N.}\ \bibnamefont {Cooper}},
  \ and\ \bibinfo {author} {\bibfnamefont {J.~R.}\ \bibnamefont {Schrieffer}},\
  }\href@noop {} {\bibfield  {journal} {\bibinfo  {journal} {Phys. Rev.}\
  }\textbf {\bibinfo {volume} {106}},\ \bibinfo {pages} {162} (\bibinfo {year}
  {1957}{\natexlab{a}})}\BibitemShut {NoStop}%
\bibitem [{\citenamefont {Bardeen}, \citenamefont {Cooper},\ and\ \citenamefont
  {Schrieffer}(1957{\natexlab{b}})}]{ref:Bar57b}%
  \BibitemOpen
  \bibfield  {author} {\bibinfo {author} {\bibfnamefont {J.}~\bibnamefont
  {Bardeen}}, \bibinfo {author} {\bibfnamefont {L.~N.}\ \bibnamefont {Cooper}},
  \ and\ \bibinfo {author} {\bibfnamefont {J.~R.}\ \bibnamefont {Schrieffer}},\
  }\href@noop {} {\bibfield  {journal} {\bibinfo  {journal} {Phys. Rev.}\
  }\textbf {\bibinfo {volume} {108}},\ \bibinfo {pages} {1175} (\bibinfo {year}
  {1957}{\natexlab{b}})}\BibitemShut {NoStop}%
\bibitem [{\citenamefont {Racah}(1943)}]{ref:Rac43}%
  \BibitemOpen
  \bibfield  {author} {\bibinfo {author} {\bibfnamefont {G.}~\bibnamefont
  {Racah}},\ }\href@noop {} {\bibfield  {journal} {\bibinfo  {journal} {Phys.
  Rev.}\ }\textbf {\bibinfo {volume} {63}},\ \bibinfo {pages} {367} (\bibinfo
  {year} {1943})}\BibitemShut {NoStop}%
\bibitem [{\citenamefont {Kerman}(1961)}]{ref:Ker61}%
  \BibitemOpen
  \bibfield  {author} {\bibinfo {author} {\bibfnamefont {A.~K.}\ \bibnamefont
  {Kerman}},\ }\href@noop {} {\bibfield  {journal} {\bibinfo  {journal} {Ann.
  Phys. (N. Y.)}\ }\textbf {\bibinfo {volume} {12}},\ \bibinfo {pages} {300}
  (\bibinfo {year} {1961})}\BibitemShut {NoStop}%
\bibitem [{\citenamefont {Talmi}(1993)}]{ref:Tal93}%
  \BibitemOpen
  \bibfield  {author} {\bibinfo {author} {\bibfnamefont {I.}~\bibnamefont
  {Talmi}},\ }\href@noop {} {\emph {\bibinfo {title} {Simple Models of Complex
  Nuclei}}}\ (\bibinfo  {publisher} {Harwood Academic Publishers},\ \bibinfo
  {address} {Chur, Switzerland},\ \bibinfo {year} {1993})\BibitemShut {NoStop}%
\bibitem [{\citenamefont {Cartan}(1894)}]{ref:Car94}%
  \BibitemOpen
  \bibfield  {author} {\bibinfo {author} {\bibfnamefont {E.}~\bibnamefont
  {Cartan}},\ }\href@noop {} {\emph {\bibinfo {title} {Sur la Structure des
  Groupes de Transformations Finis et Continus}}}\ (\bibinfo  {publisher}
  {Librairie Nony \& C$^\text{ie}$},\ \bibinfo {address} {Paris, France},\
  \bibinfo {year} {1894})\BibitemShut {NoStop}%
\bibitem [{\citenamefont {Racah}()}]{ref:Rac51}%
  \BibitemOpen
  \bibfield  {author} {\bibinfo {author} {\bibfnamefont {G.}~\bibnamefont
  {Racah}},\ }\href@noop {} {\emph {\bibinfo {title} {Group Theory and
  Spectroscopy}}},\ \bibinfo {note} {{ lectures, Princeton University, Spring
  1951, notes by E. Merzbacher and D. Pank avilable from the CERN Document
  Server with \url{http://cds.cern.ch/record/104181} }}\BibitemShut {NoStop}%
\bibitem [{\citenamefont {Jacobsen}(1962)}]{ref:Jac62}%
  \BibitemOpen
  \bibfield  {author} {\bibinfo {author} {\bibfnamefont {N.}~\bibnamefont
  {Jacobsen}},\ }\href@noop {} {\emph {\bibinfo {title} {Lie Algebras}}}\
  (\bibinfo  {publisher} {Interscience Publishers},\ \bibinfo {address} {New
  York, USA},\ \bibinfo {year} {1962})\BibitemShut {NoStop}%
\bibitem [{\citenamefont {Flowers}(1952{\natexlab{a}})}]{ref:Flo52b}%
  \BibitemOpen
  \bibfield  {author} {\bibinfo {author} {\bibfnamefont {B.~H.}\ \bibnamefont
  {Flowers}},\ }\href@noop {} {\bibfield  {journal} {\bibinfo  {journal} {Proc.
  R. Soc. Lond. A}\ }\textbf {\bibinfo {volume} {212}},\ \bibinfo {pages} {248}
  (\bibinfo {year} {1952}{\natexlab{a}})}\BibitemShut {NoStop}%
\bibitem [{\citenamefont {Jahn}(1950)}]{ref:Jah50}%
  \BibitemOpen
  \bibfield  {author} {\bibinfo {author} {\bibfnamefont {H.~A.}\ \bibnamefont
  {Jahn}},\ }\href@noop {} {\bibfield  {journal} {\bibinfo  {journal} {Proc. R.
  Soc. Lond. A}\ }\textbf {\bibinfo {volume} {201}},\ \bibinfo {pages} {516}
  (\bibinfo {year} {1950})}\BibitemShut {NoStop}%
\bibitem [{\citenamefont {Flowers}(1952{\natexlab{b}})}]{ref:Flo52a}%
  \BibitemOpen
  \bibfield  {author} {\bibinfo {author} {\bibfnamefont {B.~H.}\ \bibnamefont
  {Flowers}},\ }\href@noop {} {\bibfield  {journal} {\bibinfo  {journal} {Proc.
  R. Soc. Lond. A}\ }\textbf {\bibinfo {volume} {210}},\ \bibinfo {pages} {497}
  (\bibinfo {year} {1952}{\natexlab{b}})}\BibitemShut {NoStop}%
\bibitem [{\citenamefont {Flowers}\ and\ \citenamefont
  {Szpikowski}(1964{\natexlab{a}})}]{ref:Flo64a}%
  \BibitemOpen
  \bibfield  {author} {\bibinfo {author} {\bibfnamefont {B.~H.}\ \bibnamefont
  {Flowers}}\ and\ \bibinfo {author} {\bibfnamefont {S.}~\bibnamefont
  {Szpikowski}},\ }\href@noop {} {\bibfield  {journal} {\bibinfo  {journal}
  {Proc. Phys. Soc.}\ }\textbf {\bibinfo {volume} {84}},\ \bibinfo {pages}
  {193} (\bibinfo {year} {1964}{\natexlab{a}})}\BibitemShut {NoStop}%
\bibitem [{\citenamefont {Flowers}\ and\ \citenamefont
  {Szpikowski}(1964{\natexlab{b}})}]{ref:Flo64b}%
  \BibitemOpen
  \bibfield  {author} {\bibinfo {author} {\bibfnamefont {B.~H.}\ \bibnamefont
  {Flowers}}\ and\ \bibinfo {author} {\bibfnamefont {S.}~\bibnamefont
  {Szpikowski}},\ }\href@noop {} {\bibfield  {journal} {\bibinfo  {journal}
  {Proc. Phys. Soc.}\ }\textbf {\bibinfo {volume} {84}},\ \bibinfo {pages}
  {673} (\bibinfo {year} {1964}{\natexlab{b}})}\BibitemShut {NoStop}%
\bibitem [{\citenamefont {Edmonds}\ and\ \citenamefont
  {Flowers}(1952)}]{ref:Edm52}%
  \BibitemOpen
  \bibfield  {author} {\bibinfo {author} {\bibfnamefont {A.~R.}\ \bibnamefont
  {Edmonds}}\ and\ \bibinfo {author} {\bibfnamefont {B.~H.}\ \bibnamefont
  {Flowers}},\ }\href@noop {} {\bibfield  {journal} {\bibinfo  {journal} {Proc.
  R. Soc. Lond. A}\ }\textbf {\bibinfo {volume} {214}},\ \bibinfo {pages} {515}
  (\bibinfo {year} {1952})}\BibitemShut {NoStop}%
\bibitem [{\citenamefont {Bayman}(1960)}]{ref:Bay60}%
  \BibitemOpen
  \bibfield  {author} {\bibinfo {author} {\bibfnamefont {B.~F.}\ \bibnamefont
  {Bayman}},\ }\href@noop {} {\emph {\bibinfo {title} {Some Lectures on Groups
  and their Applications to Spectroscopy}}}\ (\bibinfo  {publisher} {Nordita},\
  \bibinfo {address} {Copenhagen, Denmark},\ \bibinfo {year}
  {1960})\BibitemShut {NoStop}%
\bibitem [{\citenamefont {Engel}, \citenamefont {Langanke},\ and\ \citenamefont
  {Vogel}(1996)}]{ref:Eng96}%
  \BibitemOpen
  \bibfield  {author} {\bibinfo {author} {\bibfnamefont {J.}~\bibnamefont
  {Engel}}, \bibinfo {author} {\bibfnamefont {K.}~\bibnamefont {Langanke}}, \
  and\ \bibinfo {author} {\bibfnamefont {P.}~\bibnamefont {Vogel}},\
  }\href@noop {} {\bibfield  {journal} {\bibinfo  {journal} {Phys. Lett. B}\
  }\textbf {\bibinfo {volume} {389}},\ \bibinfo {pages} {211} (\bibinfo {year}
  {1996})}\BibitemShut {NoStop}%
\bibitem [{\citenamefont {Engel}\ \emph {et~al.}(1997)\citenamefont {Engel},
  \citenamefont {Pittel}, \citenamefont {Stoitsov}, \citenamefont {Vogel},\
  and\ \citenamefont {Dukelsky}}]{ref:Eng97}%
  \BibitemOpen
  \bibfield  {author} {\bibinfo {author} {\bibfnamefont {J.}~\bibnamefont
  {Engel}}, \bibinfo {author} {\bibfnamefont {S.}~\bibnamefont {Pittel}},
  \bibinfo {author} {\bibfnamefont {M.}~\bibnamefont {Stoitsov}}, \bibinfo
  {author} {\bibfnamefont {P.}~\bibnamefont {Vogel}}, \ and\ \bibinfo {author}
  {\bibfnamefont {J.}~\bibnamefont {Dukelsky}},\ }\href@noop {} {\bibfield
  {journal} {\bibinfo  {journal} {Phys. Rev. C}\ }\textbf {\bibinfo {volume}
  {55}},\ \bibinfo {pages} {1781} (\bibinfo {year} {1997})}\BibitemShut
  {NoStop}%
\bibitem [{\citenamefont {Kota}\ and\ \citenamefont {{Castilho
  Alcar\'as}}(2006)}]{ref:Kot06}%
  \BibitemOpen
  \bibfield  {author} {\bibinfo {author} {\bibfnamefont {V.~K.~B.}\
  \bibnamefont {Kota}}\ and\ \bibinfo {author} {\bibfnamefont {J.~A.}\
  \bibnamefont {{Castilho Alcar\'as}}},\ }\href@noop {} {\bibfield  {journal}
  {\bibinfo  {journal} {Nucl. Phys. A}\ }\textbf {\bibinfo {volume} {764}},\
  \bibinfo {pages} {181} (\bibinfo {year} {2006})}\BibitemShut {NoStop}%
\bibitem [{\citenamefont {Dukelsky}\ \emph {et~al.}(2006)\citenamefont
  {Dukelsky}, \citenamefont {Gueorguiev}, \citenamefont {{Van Isacker}},
  \citenamefont {Dimitrova}, \citenamefont {Errea},\ and\ \citenamefont {{Lerma
  H.}}}]{ref:Duk06}%
  \BibitemOpen
  \bibfield  {author} {\bibinfo {author} {\bibfnamefont {J.}~\bibnamefont
  {Dukelsky}}, \bibinfo {author} {\bibfnamefont {V.~G.}\ \bibnamefont
  {Gueorguiev}}, \bibinfo {author} {\bibfnamefont {P.}~\bibnamefont {{Van
  Isacker}}}, \bibinfo {author} {\bibfnamefont {S.}~\bibnamefont {Dimitrova}},
  \bibinfo {author} {\bibfnamefont {B.}~\bibnamefont {Errea}}, \ and\ \bibinfo
  {author} {\bibfnamefont {S.}~\bibnamefont {{Lerma H.}}},\ }\href@noop {}
  {\bibfield  {journal} {\bibinfo  {journal} {Phys. Rev. Lett.}\ }\textbf
  {\bibinfo {volume} {96}},\ \bibinfo {pages} {072503} (\bibinfo {year}
  {2006})}\BibitemShut {NoStop}%
\bibitem [{\citenamefont {{Lerma H.}}\ \emph {et~al.}(2007)\citenamefont
  {{Lerma H.}}, \citenamefont {Errea}, \citenamefont {Dukelsky},\ and\
  \citenamefont {Satu{\l}a}}]{ref:Ler07}%
  \BibitemOpen
  \bibfield  {author} {\bibinfo {author} {\bibfnamefont {S.}~\bibnamefont
  {{Lerma H.}}}, \bibinfo {author} {\bibfnamefont {B.}~\bibnamefont {Errea}},
  \bibinfo {author} {\bibfnamefont {J.}~\bibnamefont {Dukelsky}}, \ and\
  \bibinfo {author} {\bibfnamefont {W.}~\bibnamefont {Satu{\l}a}},\ }\href@noop
  {} {\bibfield  {journal} {\bibinfo  {journal} {Phys. Rev. Lett}\ }\textbf
  {\bibinfo {volume} {99}},\ \bibinfo {pages} {032501} (\bibinfo {year}
  {2007})}\BibitemShut {NoStop}%
\bibitem [{\citenamefont {Rowe}\ and\ \citenamefont
  {Carvalho}(2007)}]{ref:Row07}%
  \BibitemOpen
  \bibfield  {author} {\bibinfo {author} {\bibfnamefont {D.~J.}\ \bibnamefont
  {Rowe}}\ and\ \bibinfo {author} {\bibfnamefont {M.~J.}\ \bibnamefont
  {Carvalho}},\ }\href@noop {} {\bibfield  {journal} {\bibinfo  {journal} {J.
  Phys. A; Math. Theor.}\ }\textbf {\bibinfo {volume} {40}},\ \bibinfo {pages}
  {471} (\bibinfo {year} {2007})}\BibitemShut {NoStop}%
\bibitem [{\citenamefont {Drumev}\ and\ \citenamefont
  {Georgieva}(2013)}]{ref:Dru13}%
  \BibitemOpen
  \bibfield  {author} {\bibinfo {author} {\bibfnamefont {K.~P.}\ \bibnamefont
  {Drumev}}\ and\ \bibinfo {author} {\bibfnamefont {A.~I.}\ \bibnamefont
  {Georgieva}},\ }\href@noop {} {\bibfield  {journal} {\bibinfo  {journal}
  {Nucl. Theor.}\ }\textbf {\bibinfo {volume} {32}},\ \bibinfo {pages} {151}
  (\bibinfo {year} {2013})}\BibitemShut {NoStop}%
\bibitem [{\citenamefont {Drumev}\ and\ \citenamefont
  {Georgieva}(2014)}]{ref:Dru14}%
  \BibitemOpen
  \bibfield  {author} {\bibinfo {author} {\bibfnamefont {K.~P.}\ \bibnamefont
  {Drumev}}\ and\ \bibinfo {author} {\bibfnamefont {A.~I.}\ \bibnamefont
  {Georgieva}},\ }\href@noop {} {\bibfield  {journal} {\bibinfo  {journal}
  {Nucl. Theor.}\ }\textbf {\bibinfo {volume} {33}},\ \bibinfo {pages} {162}
  (\bibinfo {year} {2014})}\BibitemShut {NoStop}%
\bibitem [{\citenamefont {{M\'arquez Romero}}, \citenamefont {Dobaczewski},\
  and\ \citenamefont {Pastore}(2018)}]{ref:Mar18}%
  \BibitemOpen
  \bibfield  {author} {\bibinfo {author} {\bibfnamefont {A.}~\bibnamefont
  {{M\'arquez Romero}}}, \bibinfo {author} {\bibfnamefont {J.}~\bibnamefont
  {Dobaczewski}}, \ and\ \bibinfo {author} {\bibfnamefont {A.}~\bibnamefont
  {Pastore}},\ }\href@noop {} {\bibfield  {journal} {\bibinfo  {journal} {Acta
  Phys. Pol. B}\ }\textbf {\bibinfo {volume} {49}},\ \bibinfo {pages} {347}
  (\bibinfo {year} {2018})}\BibitemShut {NoStop}%
\bibitem [{\citenamefont {Rowe}, \citenamefont {Repka},\ and\ \citenamefont
  {Carvalho}(2011)}]{ref:Row11}%
  \BibitemOpen
  \bibfield  {author} {\bibinfo {author} {\bibfnamefont {D.~J.}\ \bibnamefont
  {Rowe}}, \bibinfo {author} {\bibfnamefont {J.}~\bibnamefont {Repka}}, \ and\
  \bibinfo {author} {\bibfnamefont {M.~J.}\ \bibnamefont {Carvalho}},\
  }\href@noop {} {\bibfield  {journal} {\bibinfo  {journal} {J. Math. Phys.}\
  }\textbf {\bibinfo {volume} {52}},\ \bibinfo {pages} {013507} (\bibinfo
  {year} {2011})}\BibitemShut {NoStop}%
\bibitem [{\citenamefont {Weyl}(1939)}]{ref:Wey39}%
  \BibitemOpen
  \bibfield  {author} {\bibinfo {author} {\bibfnamefont {H.}~\bibnamefont
  {Weyl}},\ }\href@noop {} {\emph {\bibinfo {title} {The Classical Groups.
  Their Invariants and Representations}}}\ (\bibinfo  {publisher} {Princeton
  University Press},\ \bibinfo {address} {Princeton, USA},\ \bibinfo {year}
  {1939})\BibitemShut {NoStop}%
\bibitem [{\citenamefont {Littlewood}(1940)}]{ref:Lit40}%
  \BibitemOpen
  \bibfield  {author} {\bibinfo {author} {\bibfnamefont {D.~E.}\ \bibnamefont
  {Littlewood}},\ }\href@noop {} {\emph {\bibinfo {title} {The Theory of Group
  Characters and Matrix Representations of Groups}}}\ (\bibinfo  {publisher}
  {Clarendon Press},\ \bibinfo {address} {Oxford, UK},\ \bibinfo {year}
  {1940})\BibitemShut {NoStop}%
\bibitem [{\citenamefont {Edmonds}(1957)}]{ref:Edm57}%
  \BibitemOpen
  \bibfield  {author} {\bibinfo {author} {\bibfnamefont {A.~R.}\ \bibnamefont
  {Edmonds}},\ }\href@noop {} {\emph {\bibinfo {title} {Angular Momentum in
  Quantum Mechanics}}}\ (\bibinfo  {publisher} {Princeton University Press},\
  \bibinfo {address} {Princeton, USA},\ \bibinfo {year} {1957})\BibitemShut
  {NoStop}%
\bibitem [{\citenamefont {B\"orner}(1955)}]{ref:Boe63}%
  \BibitemOpen
  \bibfield  {author} {\bibinfo {author} {\bibfnamefont {H.}~\bibnamefont
  {B\"orner}},\ }\href@noop {} {\emph {\bibinfo {title} {Darstellungen von
  Gruppen. Mit Ber\"ucksichtigung der Bed\"urfnisse der Modernen Physik}}}\
  (\bibinfo  {publisher} {Springer Verlag},\ \bibinfo {address} {Berlin, West
  Germany},\ \bibinfo {year} {1955})\BibitemShut {NoStop}%
\bibitem [{\citenamefont {Racah}(1950)}]{ref:Rac50}%
  \BibitemOpen
  \bibfield  {author} {\bibinfo {author} {\bibfnamefont {G.}~\bibnamefont
  {Racah}},\ }\href@noop {} {\bibfield  {journal} {\bibinfo  {journal} {Lincei
  - Rend. Sc. fis. mat. \& nat.}\ }\textbf {\bibinfo {volume} {8}},\ \bibinfo
  {pages} {108} (\bibinfo {year} {1950})}\BibitemShut {NoStop}%
\bibitem [{\citenamefont {Racah}(1949)}]{ref:Rac49}%
  \BibitemOpen
  \bibfield  {author} {\bibinfo {author} {\bibfnamefont {G.}~\bibnamefont
  {Racah}},\ }\href@noop {} {\bibfield  {journal} {\bibinfo  {journal} {Phys.
  Rev.}\ }\textbf {\bibinfo {volume} {76}},\ \bibinfo {pages} {1352} (\bibinfo
  {year} {1949})}\BibitemShut {NoStop}%
\bibitem [{\citenamefont {Racah}(1952)}]{ref:Rac52}%
  \BibitemOpen
  \bibfield  {author} {\bibinfo {author} {\bibfnamefont {G.}~\bibnamefont
  {Racah}},\ }in\ \href@noop {} {\emph {\bibinfo {booktitle} {L. Farkas
  Memorial Volume}}},\ \bibinfo {editor} {edited by\ \bibinfo {editor}
  {\bibfnamefont {A.}~\bibnamefont {Farkas}}\ and\ \bibinfo {editor}
  {\bibfnamefont {E.}~\bibnamefont {Wigner}}}\ (\bibinfo {organization}
  {Research Council of Israel},\ \bibinfo {address} {Jerusalem, Israel},\
  \bibinfo {year} {1952})\ p.\ \bibinfo {pages} {294}\BibitemShut {NoStop}%
\bibitem [{\citenamefont {Condon}\ and\ \citenamefont
  {Shortley}(1935)}]{ref:Con35}%
  \BibitemOpen
  \bibfield  {author} {\bibinfo {author} {\bibfnamefont {E.~U.}\ \bibnamefont
  {Condon}}\ and\ \bibinfo {author} {\bibfnamefont {G.~H.}\ \bibnamefont
  {Shortley}},\ }\href@noop {} {\emph {\bibinfo {title} {The Theory of Atomic
  Spectra}}}\ (\bibinfo  {publisher} {Cambridge University Press},\ \bibinfo
  {address} {Cambridge, USA},\ \bibinfo {year} {1935})\BibitemShut {NoStop}%
\bibitem [{\citenamefont {Racah}(1942)}]{ref:Rac42}%
  \BibitemOpen
  \bibfield  {author} {\bibinfo {author} {\bibfnamefont {G.}~\bibnamefont
  {Racah}},\ }\href@noop {} {\bibfield  {journal} {\bibinfo  {journal} {Phys.
  Rev.}\ }\textbf {\bibinfo {volume} {62}},\ \bibinfo {pages} {438} (\bibinfo
  {year} {1942})}\BibitemShut {NoStop}%
\bibitem [{\citenamefont {Bell}(1959)}]{ref:Bel59}%
  \BibitemOpen
  \bibfield  {author} {\bibinfo {author} {\bibfnamefont {J.~S.}\ \bibnamefont
  {Bell}},\ }\href@noop {} {\bibfield  {journal} {\bibinfo  {journal} {Nucl.
  Phys.}\ }\textbf {\bibinfo {volume} {12}},\ \bibinfo {pages} {117} (\bibinfo
  {year} {1959})}\BibitemShut {NoStop}%
\bibitem [{\citenamefont {Wigner}(1937)}]{ref:Wig37}%
  \BibitemOpen
  \bibfield  {author} {\bibinfo {author} {\bibfnamefont {E.}~\bibnamefont
  {Wigner}},\ }\href@noop {} {\bibfield  {journal} {\bibinfo  {journal} {Phys.
  Rev.}\ }\textbf {\bibinfo {volume} {51}},\ \bibinfo {pages} {947} (\bibinfo
  {year} {1937})}\BibitemShut {NoStop}%
\end{thebibliography}%

\end{document}